\documentclass[times]{cpeauth}

\usepackage{moreverb}
\usepackage{multirow}

\newcommand\BibTeX{{\rmfamily B\kern-.05em \textsc{i\kern-.025em b}\kern-.08em T\kern-.1667em\lower.7ex\hbox{E}\kern-.125emX}}

\def\volumeyear{2016}

\usepackage[dvips,colorlinks,bookmarksopen,bookmarksnumbered,citecolor=red,urlcolor=red]{hyperref}

\usepackage{graphicx} 
\usepackage{subfig}

\usepackage[]{algorithm}
\usepackage{algpseudocode}

\renewcommand{\algorithmicforall}{\textbf{for each}}

\begin{document}

\runningheads{M. Cafaro et al.} {}

\title{Parallel Space Saving on Multi and Many-Core Processors}

\author{Massimo~Cafaro\affil{1,2} \corrauth \footnotemark[2], Marco~Pulimeno\affil{1}, Italo~Epicoco\affil{1,2} and Giovanni~Aloisio\affil{1,2}}

\address{
\affilnum{1}University of Salento, Lecce, Italy\break 
\affilnum{2}CMCC Foundation - Euro-Mediterranean Centre on Climate Change, Lecce, Italy
}

\corraddr{Facolt\`a di Ingegneria, Universit\`a del Salento, Via per Monteroni, 73100 Lecce Italy}

\begin{abstract} Given an array $\mathcal{A}$ of $n$ elements and a value $2 \leq k \leq n$, a frequent item or $k$-majority element is an element occurring in $\mathcal{A}$ more than $n/k$ times. The $k$-majority problem requires finding all of the $k$-majority elements. In this paper we deal with parallel shared-memory algorithms for frequent items; we present a shared-memory version of the Space Saving algorithm and we study its behavior with regard to accuracy and performance on many and multi-core processors, including the Intel Phi accelerator. We also investigate a hybrid MPI/OpenMP version against a pure MPI based version. Through extensive experimental results we prove that the MPI/OpenMP parallel version of the algorithm significantly enhances the performance of the earlier pure MPI version of the same algorithm. Results also prove that for this algorithm the Intel Phi accelerator does not introduce any improvement with respect to the Xeon octa-core processor.   	
\end{abstract}


\maketitle
\footnotetext[2]{E-mail: massimo.cafaro@unisalento.it}

\vspace{-6pt}

\section{Introduction}
\vspace{-2pt}
\label{introduction}

In this paper we deal with parallel shared-memory algorithms for frequent items.  In data mining, this problem is usually associated to two contexts, the on--line (stream) and the off--line setting, the difference being that in the former case we are restricted to a single scan of the input. In practice, this implies that verifying the frequent items that have been found in order to discard false positives is not allowed, while in the latter case a parallel scan of the input can be used to determine the actual frequent items. Finding frequent items is also referred to as \textit{hot list analysis} \cite{Gibbons} or \textit{market basket analysis} \cite{Brin}. 

In the context of data bases, the problem is usually called an \textit{iceberg query} \cite{Fang98computingiceberg}, \cite{Beyer99bottom-upcomputation}. The name arises from
the fact that the number of frequent items is often very small (the tip of an iceberg) when compared to the large amount of input data (the iceberg).

Given an array $\mathcal{A}$ of $n$ elements and a value $2 \leq k \leq n$, a frequent item or $k$-majority element is an element occurring in $\mathcal{A}$ more than $n/k$ times. The $k$-majority problem requires finding all of the $k$-majority elements. 

For $k = 2$, the problem reduces to the well known majority problem \cite{Moore81} \cite{Moore91} \cite{Fischer82}. The $k$-majority problem has been solved sequentially first by Misra and Gries \cite{Misra82}. In their paper, it is shown how to solve it in time $O(n~\log k)$. Besides being important from a theoretical perspective, algorithms for this problem are also extremely useful in practical contexts such as, for instance, in all of the cases (such as electronic voting) where a quorum of more than $n/k$ of all of the votes received is required  for a candidate to win; another good example is extracting essential characteristics of network traffic streams passing through internet routers:  the frequency estimation of internet packet streams \cite{DemaineLM02} is indeed an instance of the $k$-majority problem. 

Another example is monitoring internet packets in order to infer network congestion \cite{Estan}, \cite{Pan}. The
problem also arises in the context of the analysis of web query logs \cite{Charikar}, and is relevant in Computational Linguistics, for instance in connection with the
estimation of the frequencies of specific words in a given language \cite{CICLing}, or in all contexts where a verification of the Zipf--Mandelbrot law is required \cite{Zipf},
\cite{Mandelbrot} (theoretical linguistics, ecological field studies \cite{Mouillot}, etc.). We note that the class of applications considered here is characterized by the
condition $k = O(1)$.

In this paper, after a brief technical introduction to the algorithm, we provide experimental results obtained on one cluster node equipped with two octa-core Intel Xeon CPU E5-2630 v3 at 2.4 Ghz, kindly provided by CINECA in Italy. Moreover, we also take into account the Intel MIC (Many Integrated Cores) architecture as a target, and provide experimental results on the Intel Phi 7120P accelerator.

Finally, we also implemented a hybrid parallel version of our algorithm exploiting MPI (Message-Passing Interface) inter-node and OpenMP intra-node, and compare this version against a pure MPI implementation on up to 512 cores.

The rest of the paper is organized as follows. In Section~\ref{related} we give an overview of related work, in Section~\ref{algorithm}, we recall our algorithm and describe its shared-memory implementation. Next, we provide extensive experimental results in Section~\ref{benchmark}, and conclude the paper in Section~\ref{conclusions}. 

\section{Related work}
\label{related}
The $k$-majority problem has been solved sequentially first by Misra and Gries \cite{Misra82}. Demaine et al. \cite{DemaineLM02} and Karp et al. \cite{Karp} proposed independently optimal algorithms which, however, are identical to the Misra and Gries algorithm. \emph{Frequent}, the algorithm designed by Demaine et al. exploits better data structures (a doubly linked list of groups, supporting decrementing a set of counters at once in $O(1)$ time) and achieves a worst-case complexity of $O(n)$. The algorithm devised by Karp et al. is based on hashing and therefore achieves the $O(n)$ bound on average.

Cormode and Hadjieleftheriou present in  \cite{Cormode} a survey of existing algorithms, which are classified as  \emph{counter} or \emph{sketch} based. All of the algorithms we have discussed so far belong to the former class; notable examples of counters--based algorithms developed recently include \emph{LossyCounting} \cite{Manku02approximatefrequency} and \emph{Space Saving} \cite{Metwally2005} \cite{Metwally2006}. Among the sketch--based ones, we recall here \emph{CountSketch} \cite{Charikar} and \emph{CountMin} \cite{Cormode05}.

The problem of merging two data summaries naturally arises in a distributed or parallel setting, in which a data set is partitioned between two or among several data sets. The goal in this context is to merge two data summaries into a single summary which provides candidate frequent items for the union of the input data sets. In particular, in order for the merged summary to be useful, it is required that its size and error bounds are those of the input data summaries. A few years ago we designed an algorithm \cite{Cafaro-Tempesta}  \cite{Cafaro-Pulimeno} for merging in parallel counter-based data summaries which are the output of the \emph{Frequent} \cite{DemaineLM02} algorithm. 

Recently, we have designed and implemented in MPI a Parallel Space Saving algorithm \cite{CPT} for message-passing architectures. The availability of the latest Intel compilers (2017 release, v17), supporting the OpenMP v4.x specification, led us to implement a corresponding shared-memory version based on OpenMP v4. In particular, we exploit the OpenMP v4.x specification, which introduces a new user's defined reduction feature, allowing for both faster and easier porting of our previous message-passing based algorithm in the context of shared-memory architectures.

\section{The Parallel Algorithm}
\vspace{-2pt}
\label{algorithm}

In this Section, we provide implementation details regarding the parallel algorithms we have implemented. Before delving into details, we briefly recall here the definition of frequent item, which is based on a parameter $k$. 

Given an input \emph{array} $\mathcal{A}$, with $size(\mathcal{A}) = n$, a $k$--majority element (or \emph{frequent item}) is an element $x$ whose \emph{frequency} $f_\mathcal{A}(x)$ (i.e., number of occurrences of $x$ in the array $\mathcal{A}$)  is such that  $f_\mathcal{A}(x) \geq \left\lfloor {\frac{n}{k}} \right\rfloor+1$.

We recall here a few basic facts related to the sequential Space Saving algorithm. The algorithm uses exactly $k$ counters in order to solve the $k$-majority problem sequentially, and allows estimating the maximum error committed when computing the frequency of an item. Let denote by $\mathcal{S}[i].e$ and $\mathcal{S}[i].\hat f$ respectively the item $e$ and its estimated frequency $\hat f$ monitored by the $i$th counter of the stream summary $S$. When processing an item which is already monitored by a counter, its estimated frequency is incremented by one. When processing an item which is not already monitored by one of the available counters, there are two possibilities. If a counter is available, it will be in charge of monitoring the item and its estimated frequency is set to one. Otherwise, if all of the counters are already occupied, the counter storing the item with minimum frequency is incremented by one. Then the monitored item is evicted from the counter and replaced by the new item.

The parallel version of the Space Saving algorithm exploits a data parallel approach by distributing a block of the input stream to each process. The parallel algorithm works as follows. In the initial domain decomposition, each process determines the indices of the first and last element related to its block, by applying a simple block distribution, in which each process is responsible for either $\left\lfloor {n/p} \right\rfloor$ or $\left\lceil {n/p} \right\rceil$ elements, where $p$ is the number of available processes. Then, each process determines a local stream summary data structure storing its local candidates and their corresponding estimated frequencies, by using the well-known sequential algorithm, shown in the pseudocode of Algorithm~\ref{pss} as the \textit{SpaceSaving} function call. An hash table is then built, storing the local candidates as keys and their corresponding estimated frequencies as values. This hash table is then sorted in ascending order by frequencies and used as input for the parallel reduction, whose purpose is to determine global candidates. This step is carried out by means of the \textit{ParallelReduction} function. When the reduction completes, the resulting summary containing the global candidates is pruned (by removing all of the items below the threshold required for an item to be frequent) and returned. 

\begin{algorithm}
\begin{algorithmic}[1]
\Require $\mathcal{N}$, the input data stream array; $n$, the length of $\mathcal{N}$; $p$, the number of processes; $k$, the $k$-majority parameter
\Ensure an hash table containing $k$--majority candidate elements
\Procedure {ParallelSpaceSaving}{$\mathcal{N},  n, p, k$}
\State $r \leftarrow$  \Call{GetProcessRank}{}
\State $left \leftarrow \left\lfloor {r~n/p} \right\rfloor$
\Comment the index of the first element of the sub-array
\State $right \leftarrow \left\lfloor {(r+1)~n/p} \right\rfloor  - 1$
\Comment the index of the last element of the sub-array
\State $local \leftarrow  \Call{SpaceSaving}{\mathcal{N}, left, right, k}$
\Comment{determine local candidates}
\State sort $local$ by counters' frequency in ascending order
\State $global \leftarrow \Call{ParallelReduction}{local, k, COMBINE}$

\If{$r == 0$} \Comment{the process with rank 0 holds the final result of the reduction}
	\State $result \leftarrow \Call{Pruned}{global, n, k}$ 
\EndIf
	\State \Return $result$ 

\EndProcedure
\caption{Parallel Space Saving.}
\label{pss}
\end{algorithmic}
\end{algorithm}

The parallel reduction uses our user's defined \textit{combine} operator to merge two stream summaries. The \textit{combine} operator, shown as Algorithm~\ref{combine}, works as follow. We determine $m_1$ and $m_2$, which are the minimum of all of the frequencies stored respectively in $\mathcal{S}_1$ and $\mathcal{S}_2$. Since the hash tables are ordered, $m_1$ and $m_2$ are simply the frequencies stored in the first counter of $\mathcal{S}_1$ and $\mathcal{S}_2$. Then, we scan the first stream summary, and for each item in $\mathcal{S}_1$ we check if the item also appears in $\mathcal{S}_2$ by calling the \textit{find} function. In this case, we insert the entry for the item in $\mathcal{S}_C$, storing as its estimated frequency the sum of its frequencies appearing in $\mathcal{S}_1$ and $\mathcal{S}_2$. Otherwise, if the current item does not belong to $\mathcal{S}_2$, we insert the entry for the item storing as its estimated frequency the sum of its frequency and the minimum of all of the frequencies in $\mathcal{S}_2$.

We then scan the second stream summary $\mathcal{S}_2$. Since each time an item in $\mathcal{S}_1$ which was also present in $\mathcal{S}_2$ has been removed from $\mathcal{S}_2$, now $\mathcal{S}_2$ contains only items that do not appear in $\mathcal{S}_1$. For each item in $\mathcal{S}_2$ we simply insert the item in $\mathcal{S}_C$ and we store as its estimated frequency the sum of its frequency and the minimum frequencies of all of the items in $\mathcal{S}_1$. Finally, the entries in $\mathcal{S}_C$ are sorted by the frequencies. Note that the
combined stream summaries $\mathcal{S}_C$ contain exactly $2k$ counters which may monitor at most $2k$ distinct items. However, we need to return at most $k$ items. Therefore, only the $k$ counters with the greatest frequencies are kept in $\mathcal{S}_C$. In \cite{CPT} we proved that the correctness and the error bounds are preserved by the parallel reduction.

Regarding the shared-memory version, we used OpenMP v4. The input dataset, an array of $n$ elements, is partitioned among $t$ OpenMP threads by using a block-based domain decomposition, in which each thread determines the indices of the first and last element related to its block, so that each thread is responsible for either $\left\lfloor {n/t} \right\rfloor$ or $\left\lceil {n/t} \right\rceil$ elements. 

After declaring the user's defined reduction, the algorithm works exactly as described in \cite{CPT} for its message-passing based counterpart, with OpenMP threads executing in a parallel region the sequential Space Saving algorithm \cite{Metwally2005} \cite{Metwally2006} on their own block of the input dataset, and producing corresponding stream summaries which are then merged together using the user's defined reduction before exiting the parallel region.

\begin{algorithm}
\begin{algorithmic}[1]
\Require $\mathcal{S}_1$, $\mathcal{S}_2$: hash tables ordered by counters' frequency; $k$, the $k$-majority parameter
\Ensure an hash table, which is the \textit{combined summary} $\mathcal{S}_C$
\Procedure {combine}{$\mathcal{S}_1$,  $\mathcal{S}_2$, $k$}
\State $m_1 \leftarrow \mathcal{S}_1[0].\hat f $
\Comment minimum of all of the frequencies in $\mathcal{S}_1$
\State $m_2 \leftarrow \mathcal{S}_2[0].\hat f $
\Comment minimum of all of the frequencies in $\mathcal{S}_2$

\State let $\mathcal{S}_C$ be an empty hash table
\ForAll{$counter_1$ in $\mathcal{S}_1$}
	\State $newcounter.item \leftarrow counter_1.item$	
	\State $counter_2 \leftarrow \mathcal{S}_2.\Call{Find}{counter_1.item}$
	\If{$counter_2$}
		\State $newcounter.\hat{f} \leftarrow counter_1.\hat{f} + counter_2.\hat{f}$
		\State $\mathcal{S}_2.\Call {Remove}{counter_2}$
	\Else
		\State $newcounter.\hat{f} \leftarrow counter_1.\hat{f} + m_2$
	\EndIf
	\State $\mathcal{S}_C.\Call {Put}{newcounter}$
\EndFor

\ForAll{$counter_2$ in $\mathcal{S}_2$}
	\State $newcounter.item \leftarrow counter_2.item$
	\State $newcounter.\hat{f} \leftarrow counter_2.\hat{f} + m_1$
	\State $\mathcal{S}_C.\Call {Put}{newcounter}$
\EndFor
\State $\mathcal{S}_C.\Call {Prune}{k}$
\Comment Select $k$ counters with the greatest frequencies and delete the others 
\State \Return $\mathcal{S}_C$
\EndProcedure
\caption{Combine.}
\label{combine}
\end{algorithmic}
\end{algorithm}

We also developed a hybrid MPI/OpenMP version, in which we take advantage of OpenMP threads intra-node, and use MPI processes inter-node. In this version, the input array is initially partitioned among the MPI processes, and then each MPI process sub-array is partitioned again among the available OpenMP threads. Once the subarray has been processed in the OpenMP parallel region, and a corresponding output stream summary has been produced at the end of the parallel region, this summary is then used as input for the MPI user's defined reduction in which the MPI processes' summaries are merged together to produce the algorithm's final output. 

Finally, for the MIC version, we used the MPI/OpenMP version in which we offload the execution of the Space Saving algorithm and the subsequent parallel user's defined reduction of the MPI processes to the Intel Phi accelerator. I/O operations are instead executed on the CPU.

\section{Experimental Results}
\vspace{-2pt}
\label{benchmark}

In this Section we report the experimental results we obtained carrying out several experiments on the Galileo cluster machine at CINECA in Italy. This machine is a linux CentOS 7.0 NeXtScale cluster with 516 compute nodes; each node is equipped with 2 2.40 GHz octa-core Intel Xeon CPUs E5-2630 v3, 128 GB RAM and 2 16 GB Intel Phi 7120P accelerators (available on 384 nodes only). High-Performance networking among the nodes is provided by Intel QDR (40Gb/s) Infiniband. All of the codes were compiled using the Intel C++ compiler v17.  

Let $f$ be the true frequency of an item and $\hat{f}$ the corresponding frequency reported by an algorithm, then the relative error is defined as $\Delta f = \frac{{\left| {f - \hat{f}} \right|}}{f}$, and the average relative error is derived by averaging the absolute relative errors over all of the measured frequencies.

Precision, a metric defined as the total number of true $k$-majority elements reported over the total number of items reported, quantifies the number of false positives reported by an algorithm in the output stream summary. Recall is the total number of true $k$-majority elements reported over the number of true $k$-majority elements given by an exact algorithm. In all of the results we obtained 100\% recall (since the algorithm is deterministic) and precision (owing to the use of the Space Saving algorithm); for this reason, to avoid wasting space, we do not show here precision plots. Rather, we present Average Relative Error (ARE) and runtime/performance plots since we are interested in understanding the error behavior and the algorithm's scalability when we use an increasing number of cores of execution. Table~\ref{experiments} summarizes all of the experiments and in the next sub-sections we present and discuss the results obtained.

\begin{table} 
 \caption{Design of experiments. The input stream size $n$ is expressed in billions, the number of Space Saving counters $k$ is expressed in thousands and $\rho$ denotes the skewness of the input data distribution.}
      \label{experiments}
	\centering
    \begin{tabular}{|c|p{2.5cm}|c|c|p{2.7cm}|}
    \hline
    Exp. & Technology & Varying & $n$, $k$, $\rho$ & Processing Elements  \\ \hline
    \multirow{3}{*}{1}  & \multirow{3}{\linewidth}{OpenMP (Xeon)} & $n=\{4, 8, 16, 29\}$ & $k=2$,  $\rho=1.1$ & \multirow{3}{\linewidth}{Thrs: 1, 2, 4, 8, 16} \\ \cline{3-4}
      &  & $k=\{0.5, 1, 2, 4, 8\}$ & $n=8$, $\rho=1.1$ & \\ \cline{3-4}
      &  & $\rho=\{1.1, 1.8\}$ & $n=8$, $k=2$ & \\ \hline \hline
     \multirow{3}{*}{2}  & \multirow{3}{\linewidth}{MPI/OpenMP vs MPI (Xeon)} & $n=\{4, 8, 16, 29\}$ & $k=2$,  $\rho=1.1$ & \multirow{3}{\linewidth}{Cores: 1, 32, 64, 128, 256, 512} \\ \cline{3-4}
      &  & $k=\{0.5, 1, 2, 4, 8\}$ & $n=29$, $\rho=1.1$ & \\ \cline{3-4}
      &  & $\rho=\{1.1, 1.8\}$ & $n=29$, $k=2$ & \\ \hline \hline
    \multirow{2}{*}{3}  & \multirow{2}{\linewidth}{OpenMP (MIC)} & $k=\{0.5, 1, 2, 4, 8\}$ & $n=3$,  $\rho=1.1$ & \multirow{2}{\linewidth}{Thrs: 15, 30, 60, 120, 240} \\ \cline{3-4} 
     &  & $\rho=\{1.1, 1.8\}$ & $n=3$,  $k=2$ &  \\ \hline \hline
    \multirow{2}{*}{4}  & \multirow{2}{\linewidth}{Xeon vs MIC (MPI/OpenMP)} & $k=\{0.5, 1, 2, 4, 8\}$ & $n=3$,  $\rho=1.1$ & \multirow{2}{\linewidth}{Sockets: 1, 4, 8, 16, 32, 64} \\ \cline{3-4} 
     &  & $\rho=\{1.1, 1.8\}$ & $n=3$,  $k=2$ &  \\ \hline
       \end{tabular}
    \end{table}

\subsection{Evaluation of the OpenMP version on the Intel Xeon}
The first experiment aimed at analyzing the pure OpenMP parallel version of the code on 1, 2, 4, 8 and 16 Xeon cores. For this experiment we used 5 different synthetic input datasets with 1, 4, 8, 16 and 29 billion items derived from a zipfian distribution with skew $1.1$ using $k=2000$ Space Saving counters. We also evaluated the scalability and ARE with different number of counters using $k=\{500, 1000, 2000, 4000, 8000\}$ and an input dataset made of 29 billion items with skew $1.1$. Finally we executed the OpenMP parallel version with two different zipfian skew values $1.1$ and $1.8$ using 2000 counters and an input dataset of 29 billion items.

\begin{figure}[hbt]
  \centering
  \begin{tabular}{ccc}
     \subfloat[Varying $k$]{
           \includegraphics[scale=0.22]{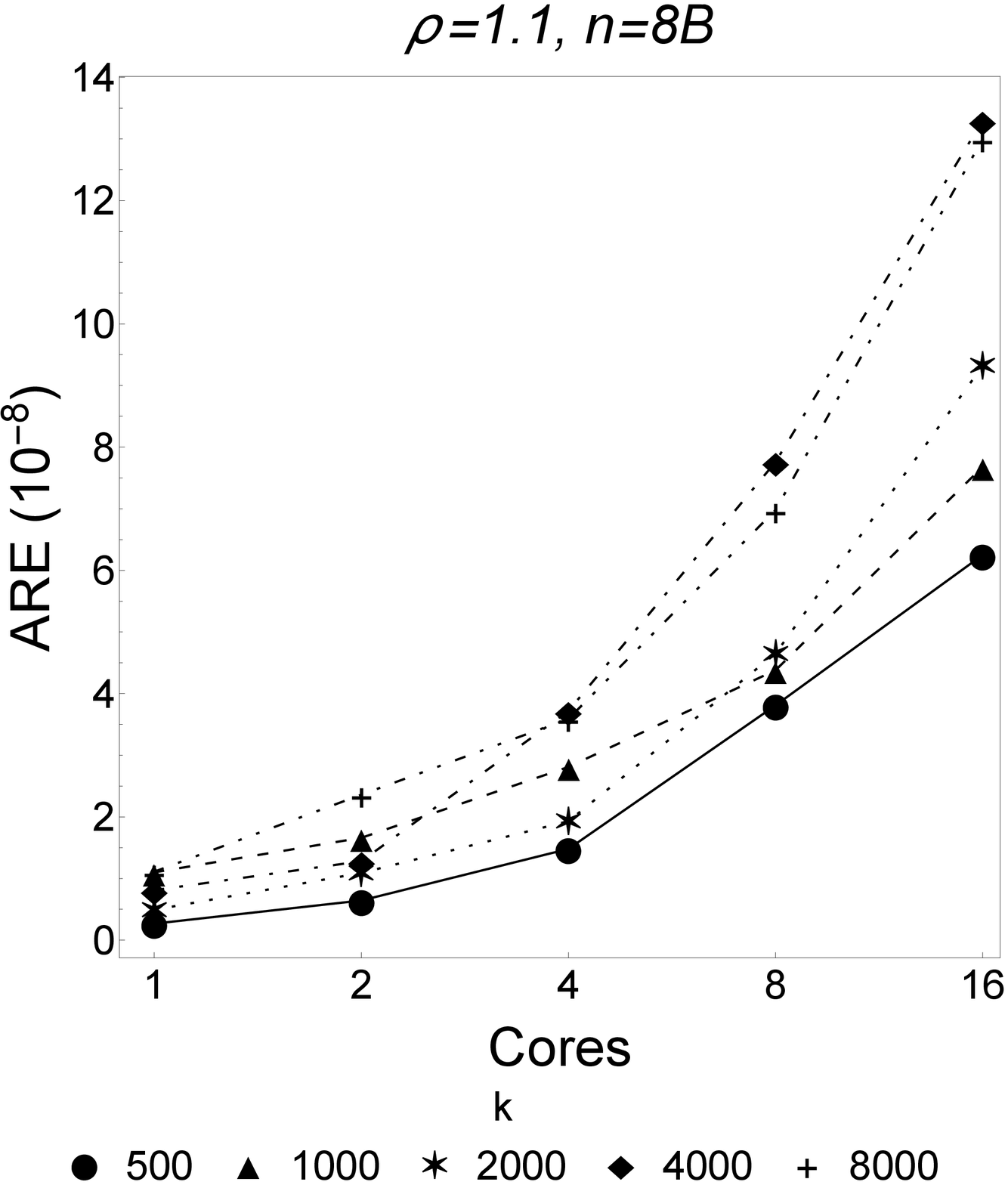}
           \label{openmp_are_k}
        } &
        
      \subfloat[Varying $n$]{
           \includegraphics[scale=0.22]{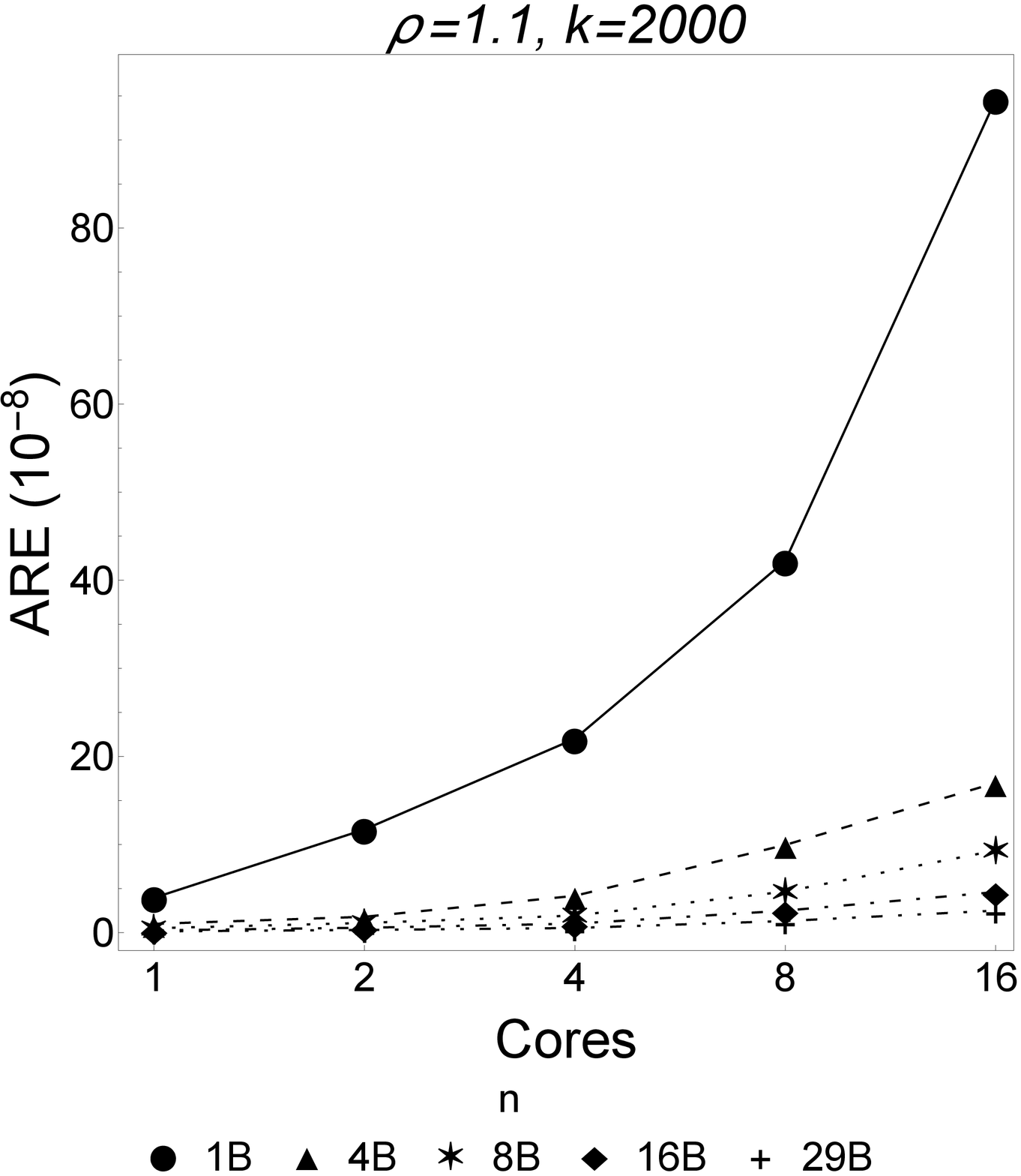}
           \label{openmp_are_n}
        } & 

      \subfloat[Varying $\rho$]{
           \includegraphics[scale=0.22]{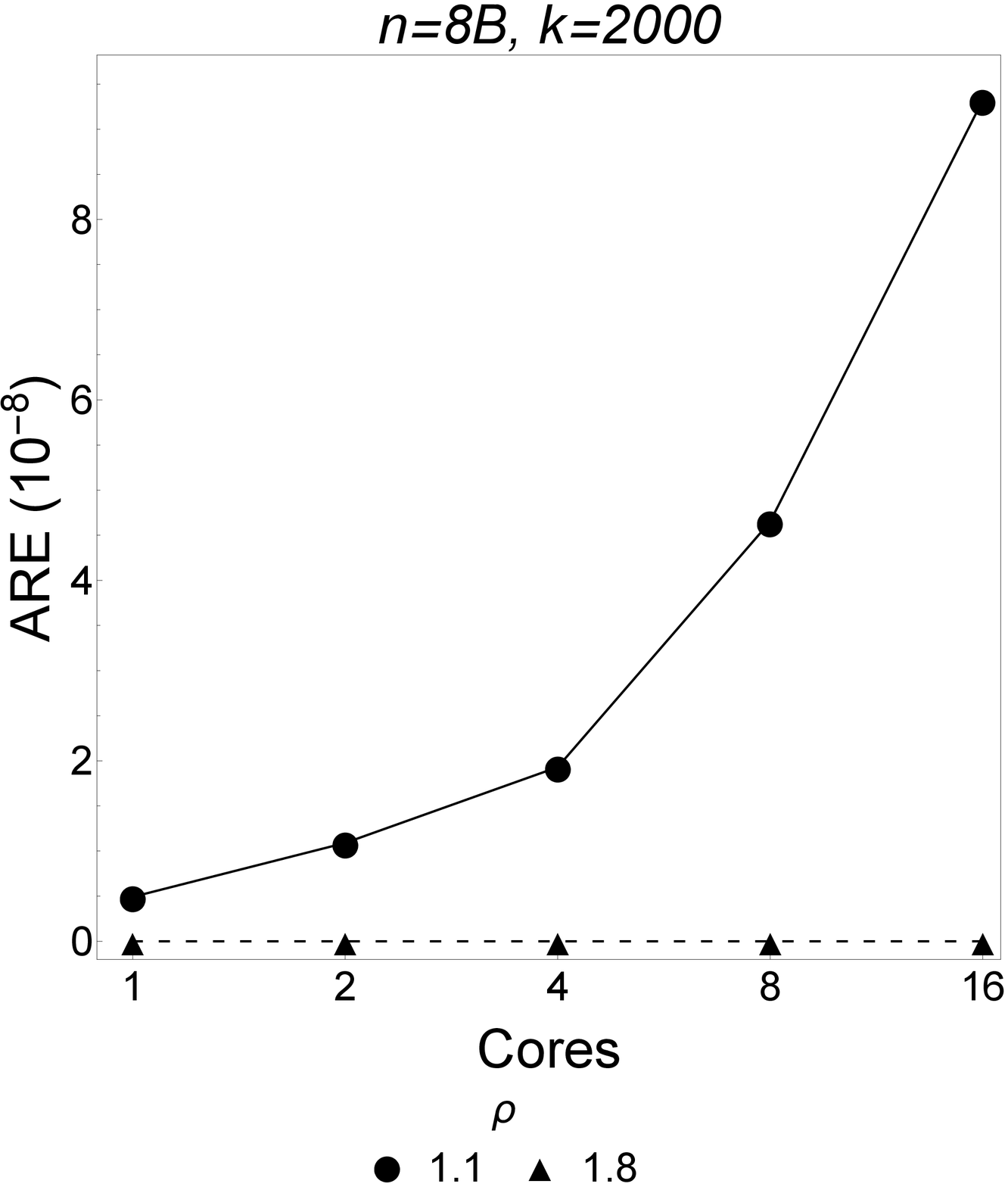}
           \label{openmp_are_ro}
        } 
\end{tabular}
 \caption{Average relative error expressed in $10^{-8}$-th of the pure OpenMP parallel version.} 
 \label{openmp_are}
\end{figure}

Figure~\ref{openmp_are} presents the Average Relative Error (ARE) we obtained in the first experiment. In particular, Figures~\ref{openmp_are_k},~\ref{openmp_are_n} and~\ref{openmp_are_ro} show the ARE values when executing the OpenMP parallel version on increasing number of cores and respectively with different numbers of Space Saving counters ($k$), different sizes of the input stream ($n$) and different skews ($\rho$) of the input distribution. In all of the cases, the ARE values are either zero or extremely low and close to zero. 

\begin{figure}[hbt]
  \centering
  \begin{tabular}{ccc}
     \subfloat[Varying $k$]{
           \includegraphics[scale=0.22]{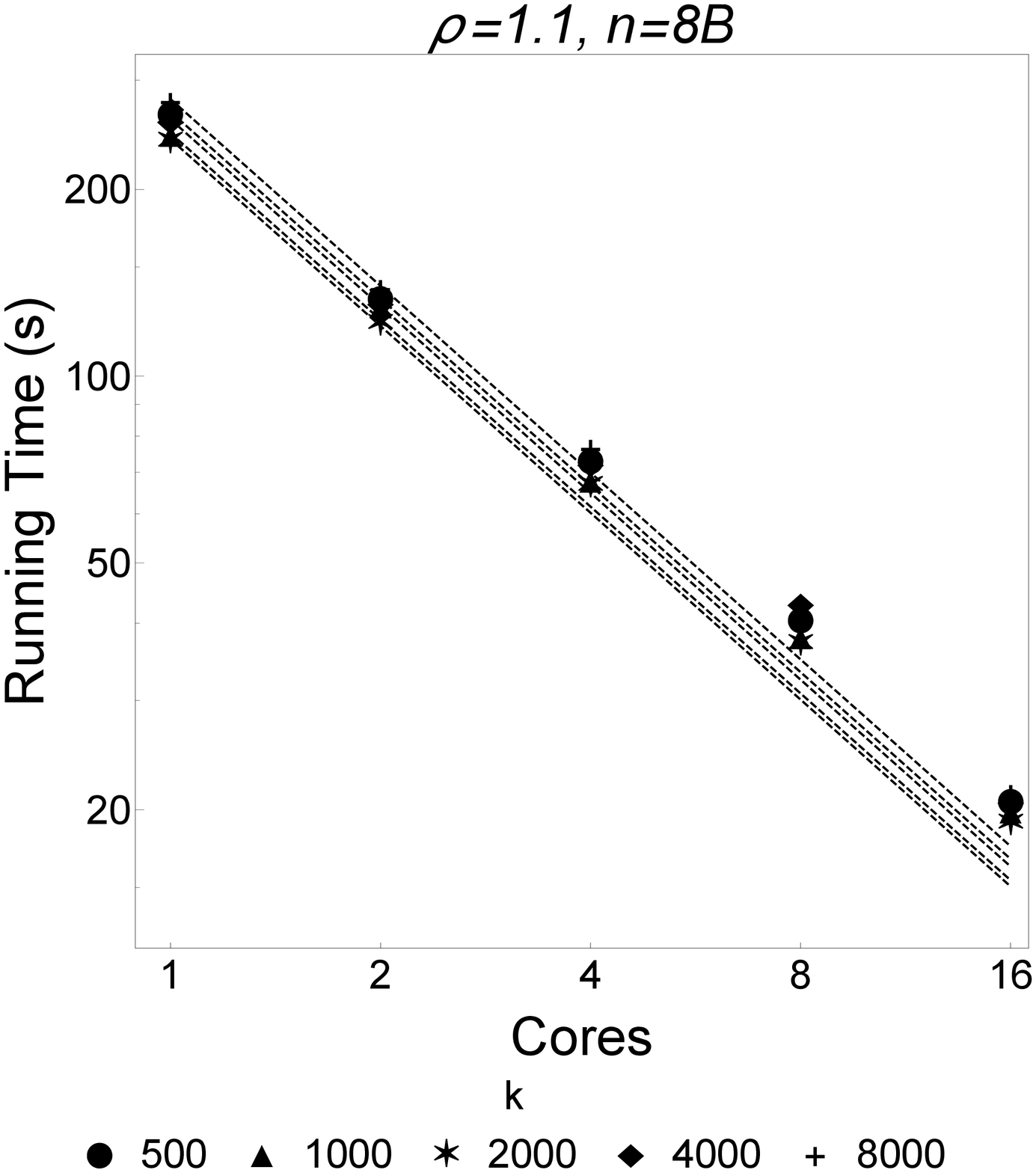}
           \label{openmp_wtime_k}
        } &
        
      \subfloat[Varying $n$]{
           \includegraphics[scale=0.22]{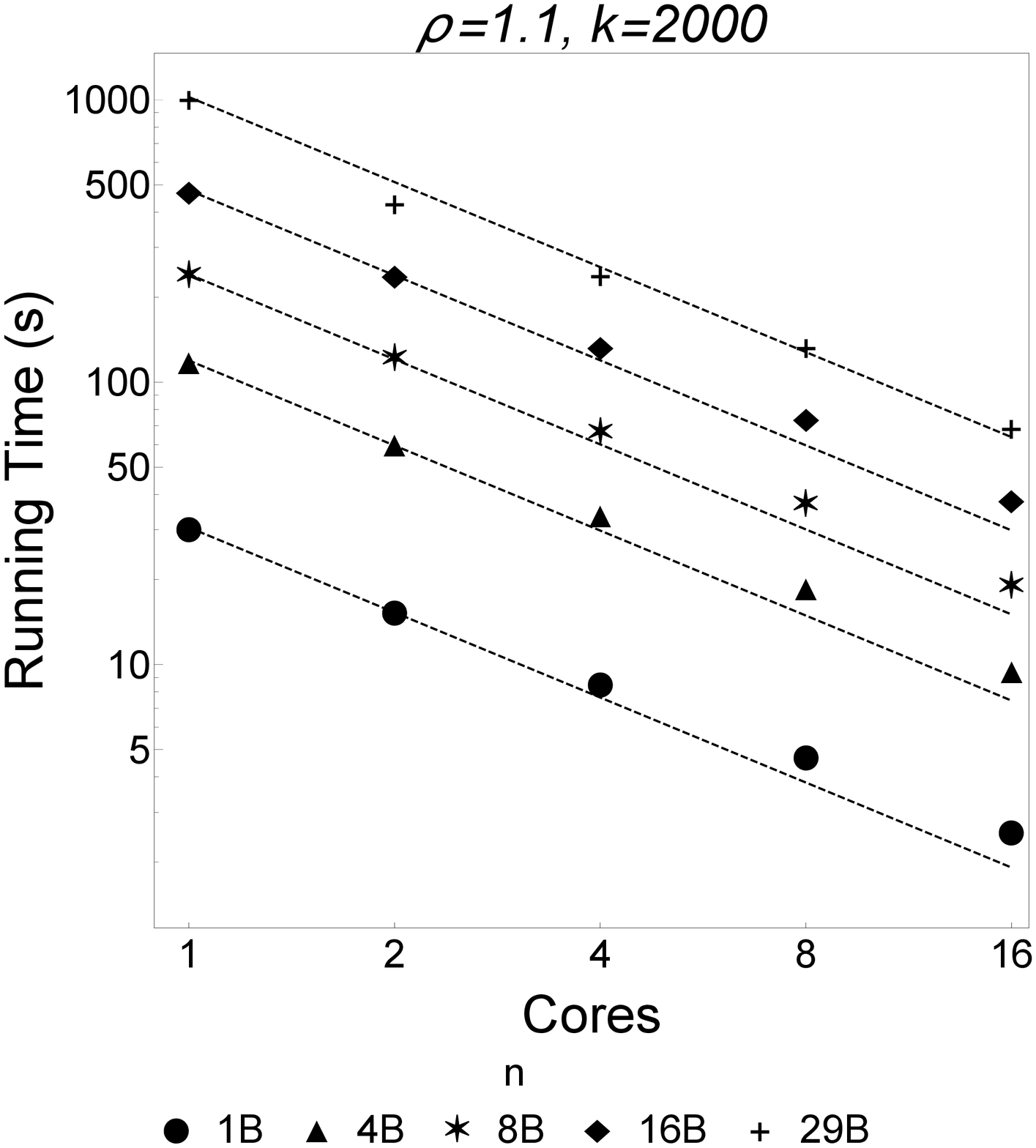}
           \label{openmp_wtime_n}
        } & 

      \subfloat[Varying $\rho$]{
           \includegraphics[scale=0.22]{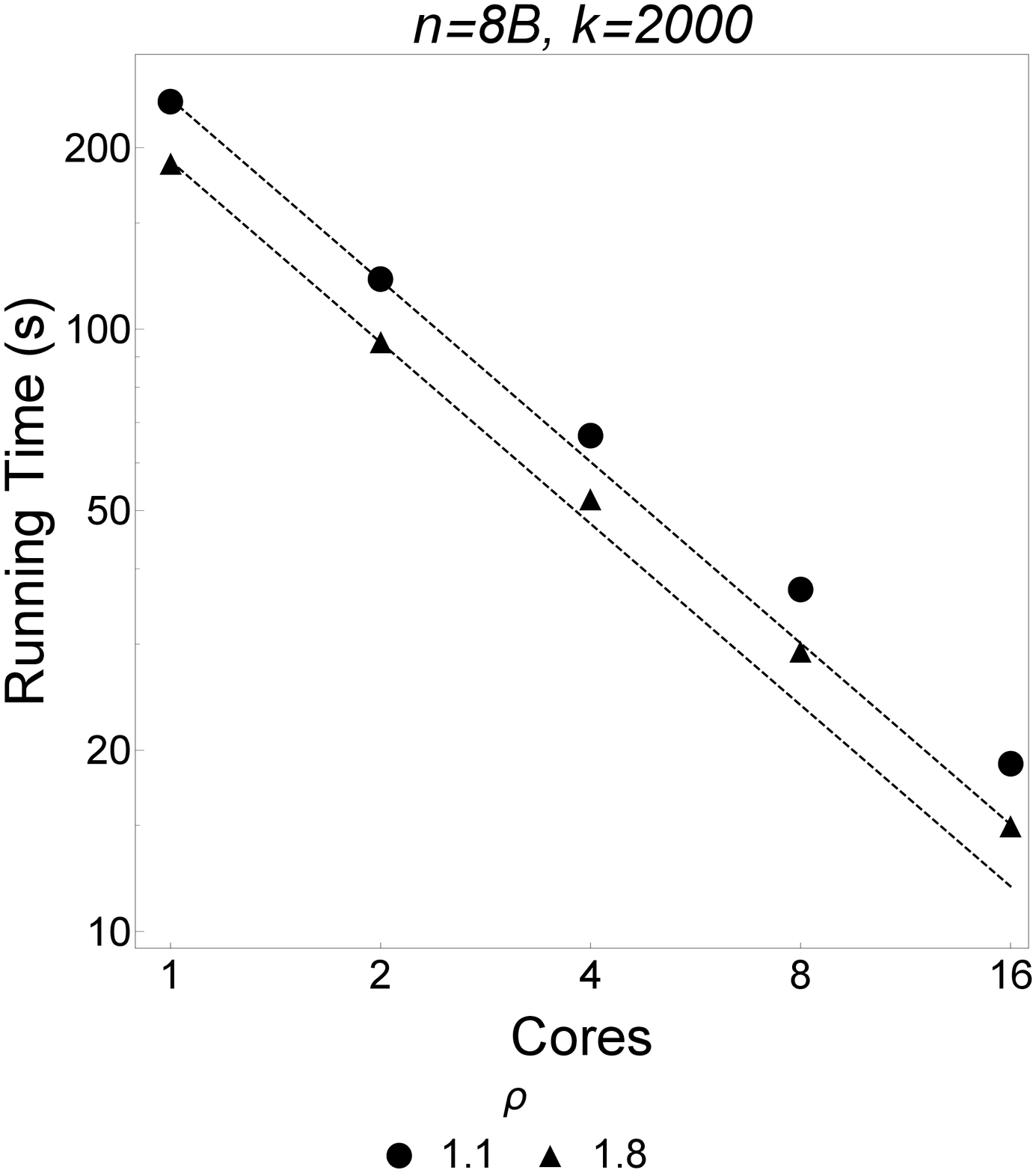}
           \label{openmp_wtime_ro}
        } 
\end{tabular} 
 \caption{Running time expressed in seconds of the pure OpenMP parallel version. The dashed lines represent ideal behavior.} 
 \label{openmp_wtime}
\end{figure}

Figure~\ref{openmp_wtime} depicts both the runtime and performance in term of scalability of the OpenMP version. Indeed, in order to analyze the scalability of the parallel algorithm, instead of plotting speedup and efficiency, we present plots using a logarithmic scale for both axes, in which we plot raw execution times varying the number of cores \cite{Heath}. For strong scaling, i.e., for a problem of fixed size, a straight line with slope -1 indicates good scalability, whereas any upward curvature away from that line indicates limited scalability. The dashed lines represent the ideal scalability for each result set. Table~\ref{openmp_wtime_data} reports the numerical results in terms of running time and speedup.

Analyzing the plots we can observe that the OpenMP parallel version reaches a good scalability. Even with small input datasets the parallel efficiency is greater than 75\% and the efficiency raises up to 92\% for bigger input stream sizes (see Figure~\ref{openmp_wtime_n}). This behaviour is expected and it is due to the so-called Amdahl effect \cite{Quinn03}, increasing the problem size the speedup increases as well. The reason is that, in general, the overhead component has lower computational complexity than the potentially parallelizable part of a computation, so increasing the problem size the complexity of the potentially parallelizable part dominates the overhead complexity and the speedup increases as well. 

\begin{figure}[hbt]
  \centering
  \begin{tabular}{cc}
     \subfloat[Varying $k$]{
           \includegraphics[scale=0.22]{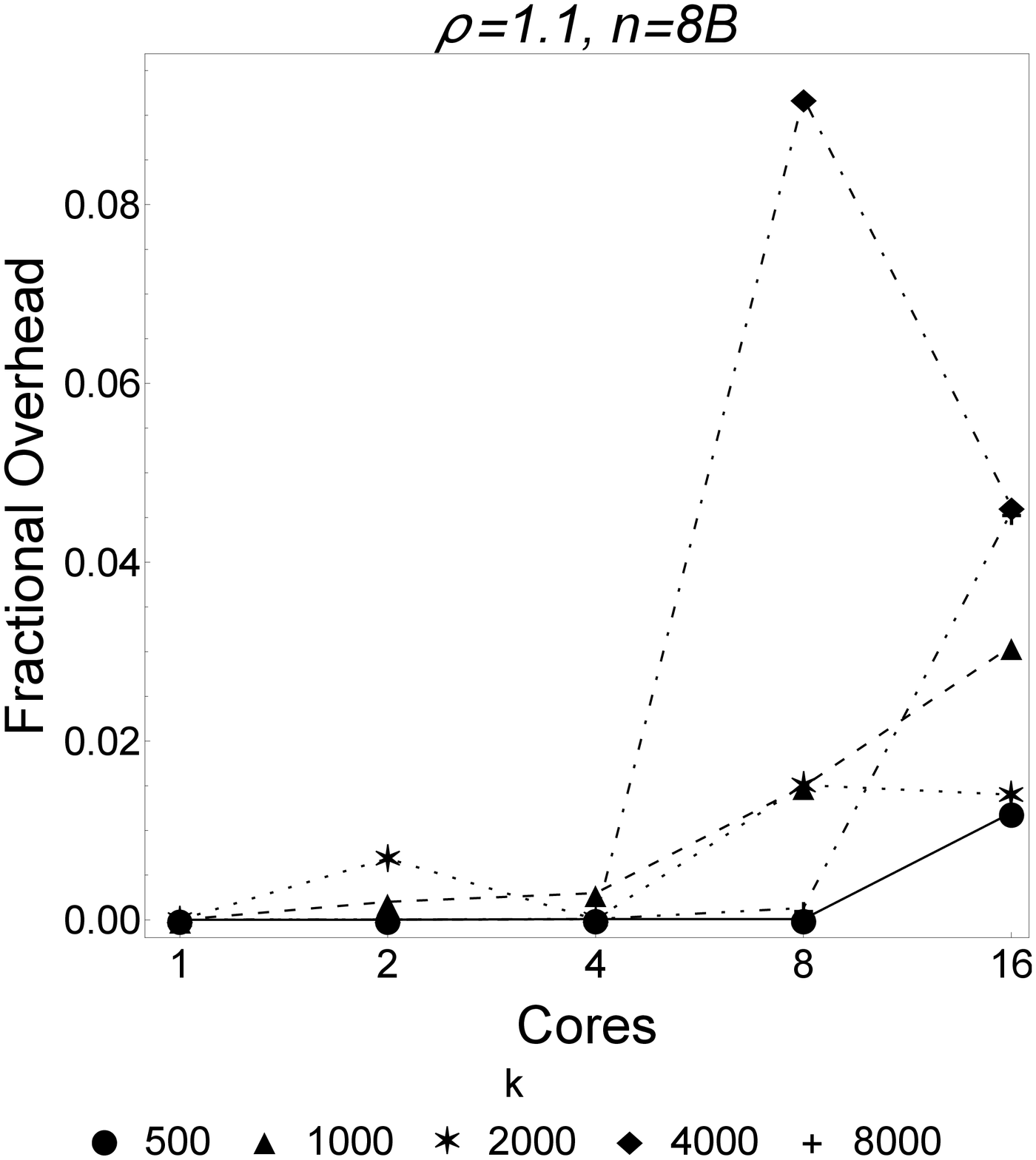}
           \label{openmp_fo_k}
        } &
        
      \subfloat[Varying $n$]{
           \includegraphics[scale=0.22]{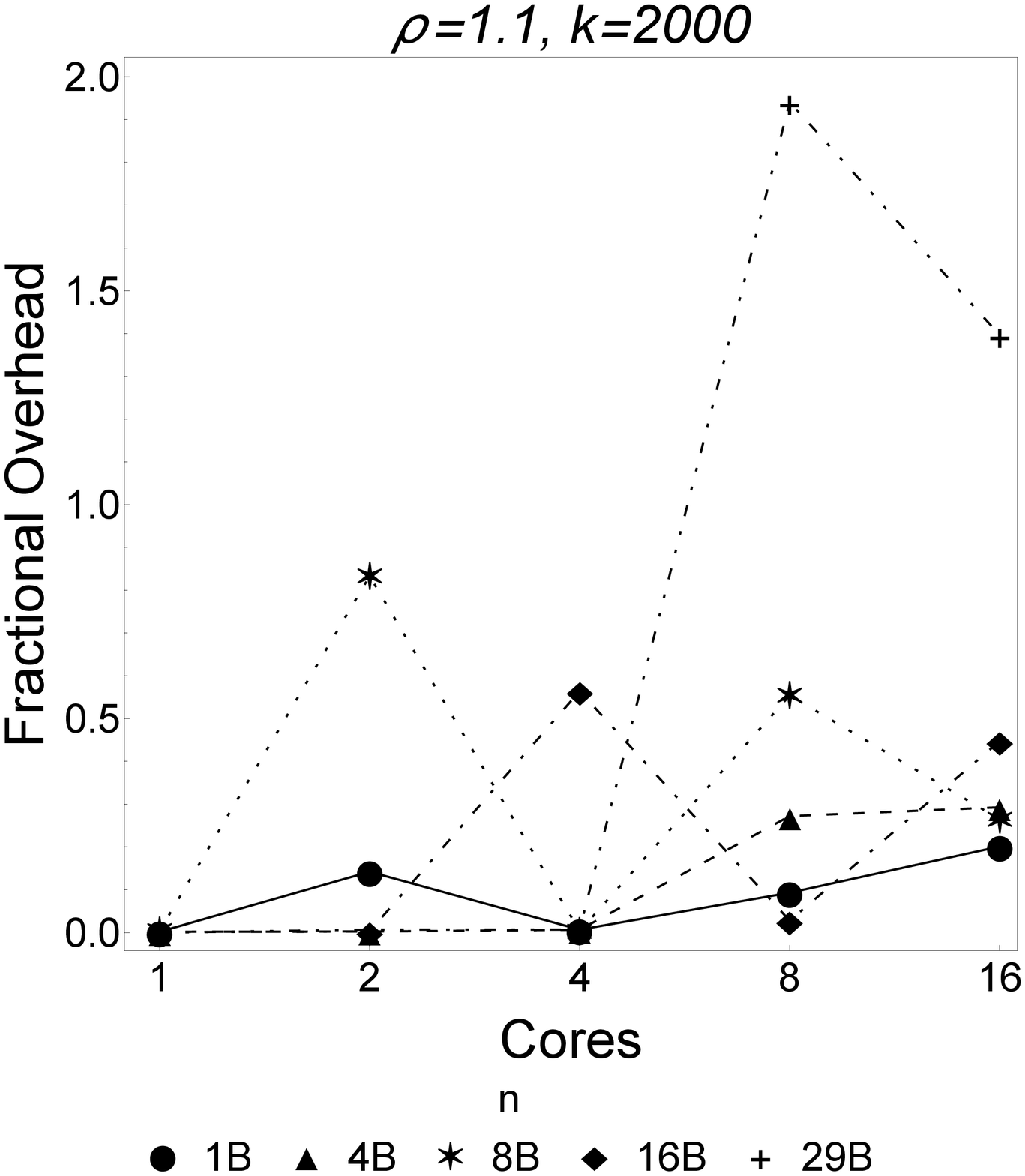}
           \label{openmp_fo_n}
        } 
\end{tabular} 
 \caption{Fractional overhead of the pure OpenMP version. The fractional overhead is the ratio of the overhead time over the computational time.} 
 \label{openmp_fo}
\end{figure}

The parallel scalability is also affected by the number of Space Saving counters used. In particular, the scalability decreases when the counters increase due to the reduction operator included in the algorithm. The greater is the number of counters, the greater is the time taken for the reduction, hence, the communication overhead grows (see Figure~\ref{openmp_wtime_k}). Finally, the skew of the input distribution does not affect significantly the parallel scalability (as reported in Figure~\ref{openmp_wtime_ro}); the input distribution skew influences only the final set of k-majority items but it does not impact on the operations computed by the algorithm. Figure~\ref{openmp_fo} depicts the fractional overhead which measures the ratio of the overhead time (that includes thread spawning, synchronization, the reduction operator) over the computational time. We measured the fractional overhead varying both the number of Space Saving counters (Figure~\ref{openmp_fo_k}) and the input stream size (Figure~\ref{openmp_fo_n}). The results show that the computational time decreases faster than the overhead when the number of threads increases, hence, the fractional overhead increases too with the number of threads and this explains why the speedup decreases.

\begin{table} [hbt]
\caption{OpenMP: in each cell the running time (in seconds) and the speedup are reported.}
\label{openmp_wtime_data}
\centering
\def\arraystretch{1.2}
{\scriptsize
\setlength\tabcolsep{5pt}
\begin{tabular}{|c|c|c|c|c|c|c|c|c|c|c|c|}
\hline
\multirow{2}{*}{Cores}  & \multicolumn{4}{c|}{Varying $n$} & \multicolumn{5}{c|}{Varying $k$} & \multicolumn{2}{c|}{Varying $\rho$} \\
 & \text{4B} & \text{8B} & \text{16B} & \text{29B}  & 500 & 1000 & 2000 & 4000 & 8000 & 1.1 & 1.8 \\
\hline
\multirow{2}{*}{1} & 120.60 & 238.45 & 481.33 & 1047.10 & 279.63 & 244.56 & 238.45 & 258.01 & 277.79 & 238.45 & 190.08 \\
 & 1 & 1 & 1 & 1 & 1 & 1 & 1 & 1 & 1 & 1 & 1 \\
\hline
\multirow{2}{*}{2} & 61.39 & 123.20 & 241.81 & 443.13 & 131.30 & 124.34 & 123.20 & 130.25 & 141.37 & 123.20 & 96.23\\
 & 1.96 & 1.94 & 1.99 & 2.36 & 2.12 & 1.96 & 1.94 & 1.98 & 1.96 & 1.94 & 1.97 \\
\hline
\multirow{2}{*}{4} & 33.84 & 69.02 & 135.80 & 247.76 & 72.41 & 69.00 & 69.02 & 70.95 & 76.24 & 69.02 & 52.79 \\
 & 3.56 & 3.45 & 3.54 & 4.22 & 3.86 & 3.54 & 3.45 & 3.63 & 3.64 & 3.45 & 3.60 \\
\hline
\multirow{2}{*}{8} & 19.15 & 38.00 & 74.82 & 138.36 & 40.54 & 38.49 & 38.00 & 39.47 & 42.13  & 38.00 & 29.47 \\
 & 6.29 & 6.28 & 6.43 & 7.56 & 6.89 & 6.35 & 6.28 & 6.53 & 6.59 & 6.28 & 6.44 \\
\hline
\multirow{2}{*}{16} & 9.74 & 19.46 & 38.77 & 71.00 & 21.24 & 19.62 & 19.46 & 20.28 & 21.72  & 19.46 & 15.11 \\
 & 12.37 & 12.25 & 12.41 & 14.74 & 13.15 & 12.45 & 12.25 & 12.71 & 12.78 & 12.25 & 12.57 \\
\hline
\end{tabular}
}
\end{table}

\subsection{Comparison between MPI and MPI/OpenMP parallel versions}
In the second experiment we analyzed and compared the performance of the pure MPI parallel version of the algorithm against the hybrid implementation based on MPI/OpenMP paradigm. Detailed results are presented in Tables~\ref{mpi_wtime_data} and~\ref{hybrid_wtime_data} where the running time and speedup are reported when varying the number of cores and input parameters such as the number of Space Saving counters, the size of the input stream and the skew of the input distribution. 

The hybrid version has been executed using 8 threads per MPI process, and each thread has been mapped to a single core. Hence, the number of cores also corresponds to the total number of threads spawned. The choice of 8 threads per process directly derives from the kind of processors available on the parallel architecture; the processors are octa-core Xeon with hyperthreading (which is the Intel implementation of SMT, Simultaneous Multi Threading) disabled.

\begin{figure}[hbt]
  \centering
  \begin{tabular}{cc}
     \subfloat[Small stream size]{
           \includegraphics[scale=0.22]{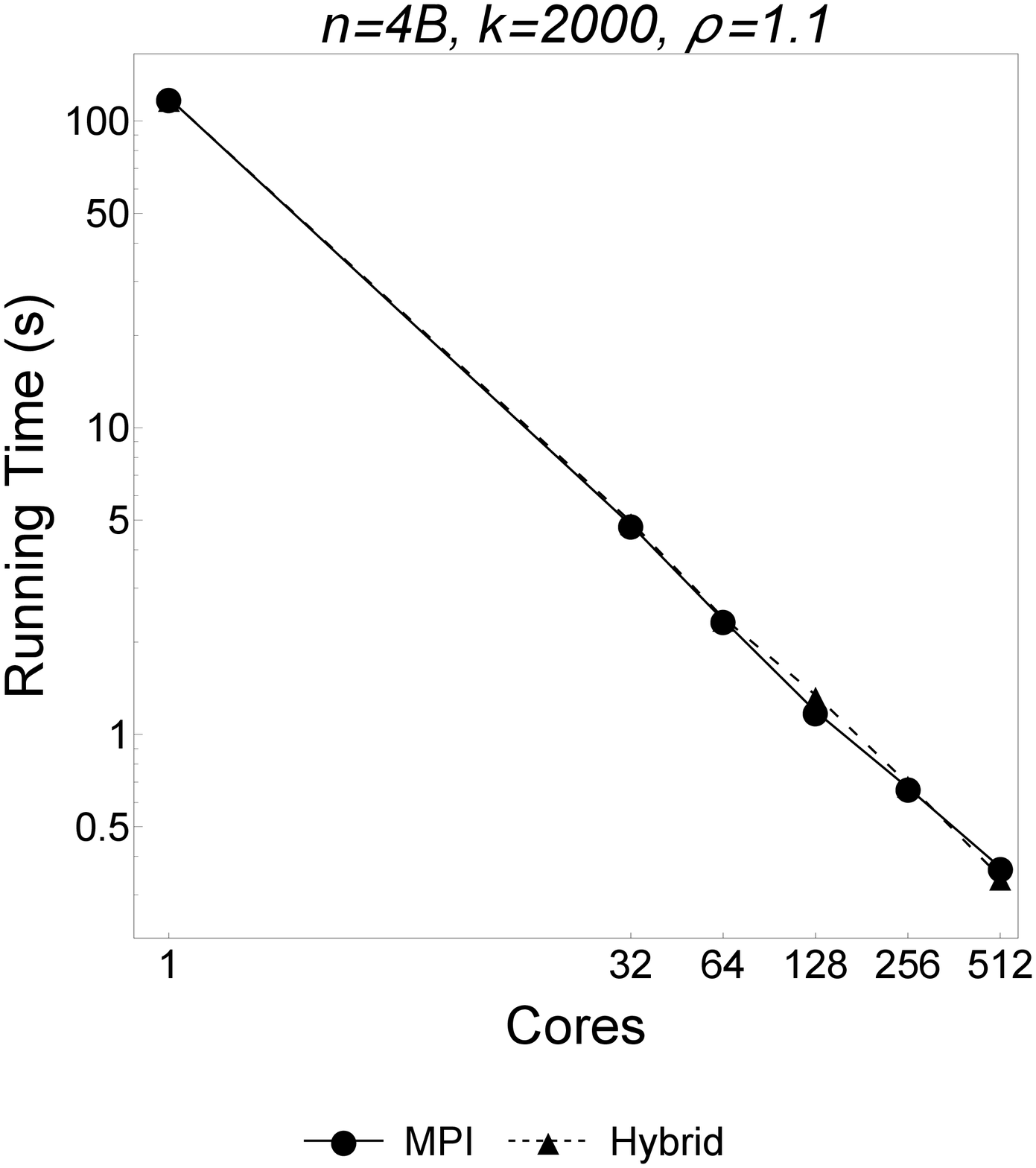}
           \label{mpi_hybrid_comparison_rt_small}
        } &

     \subfloat[Big stream size]{
           \includegraphics[scale=0.22]{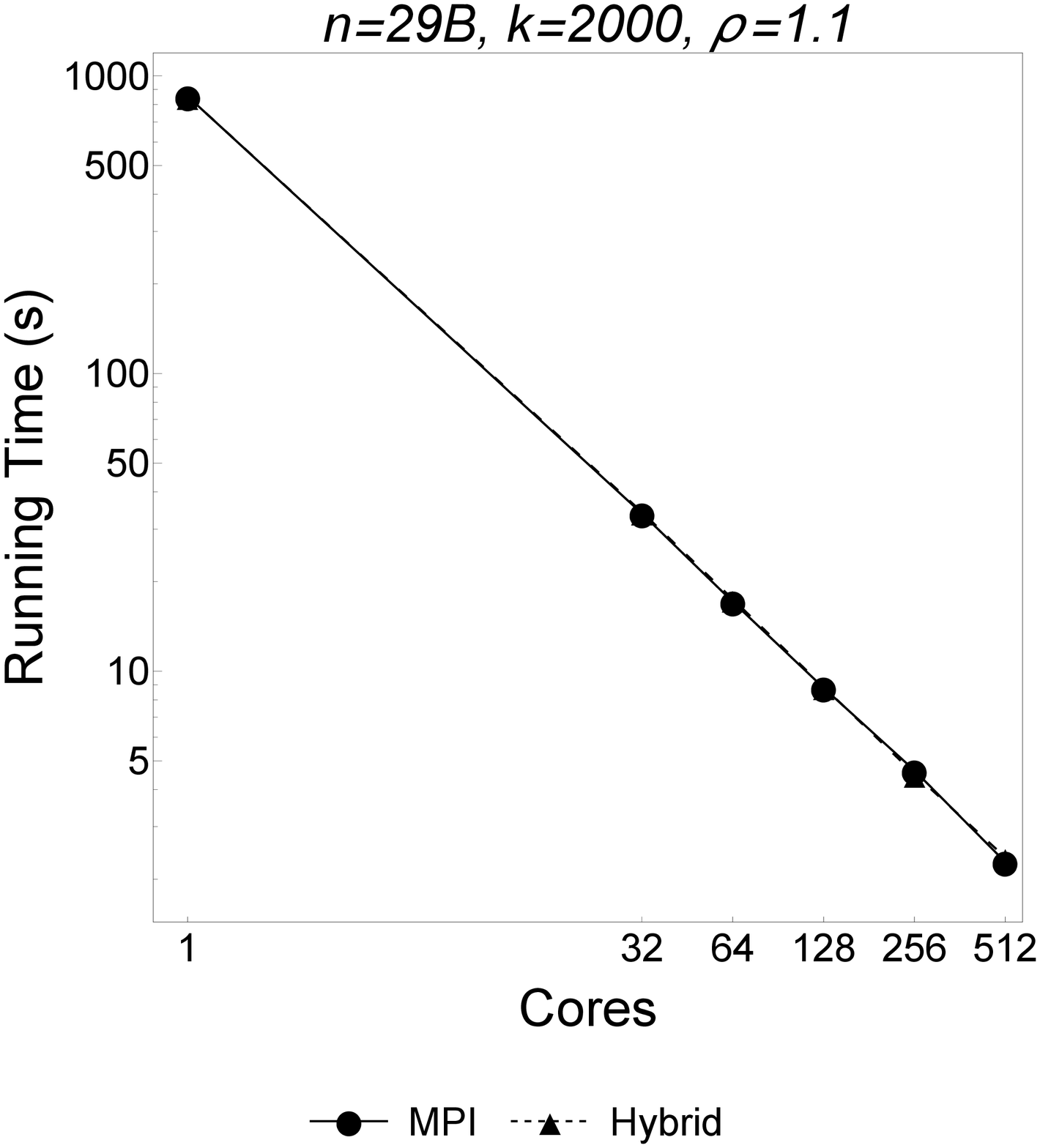}
           \label{mpi_hybrid_comparison_rt_big}
        } \\
        
     \subfloat[Small stream size]{
           \includegraphics[scale=0.22]{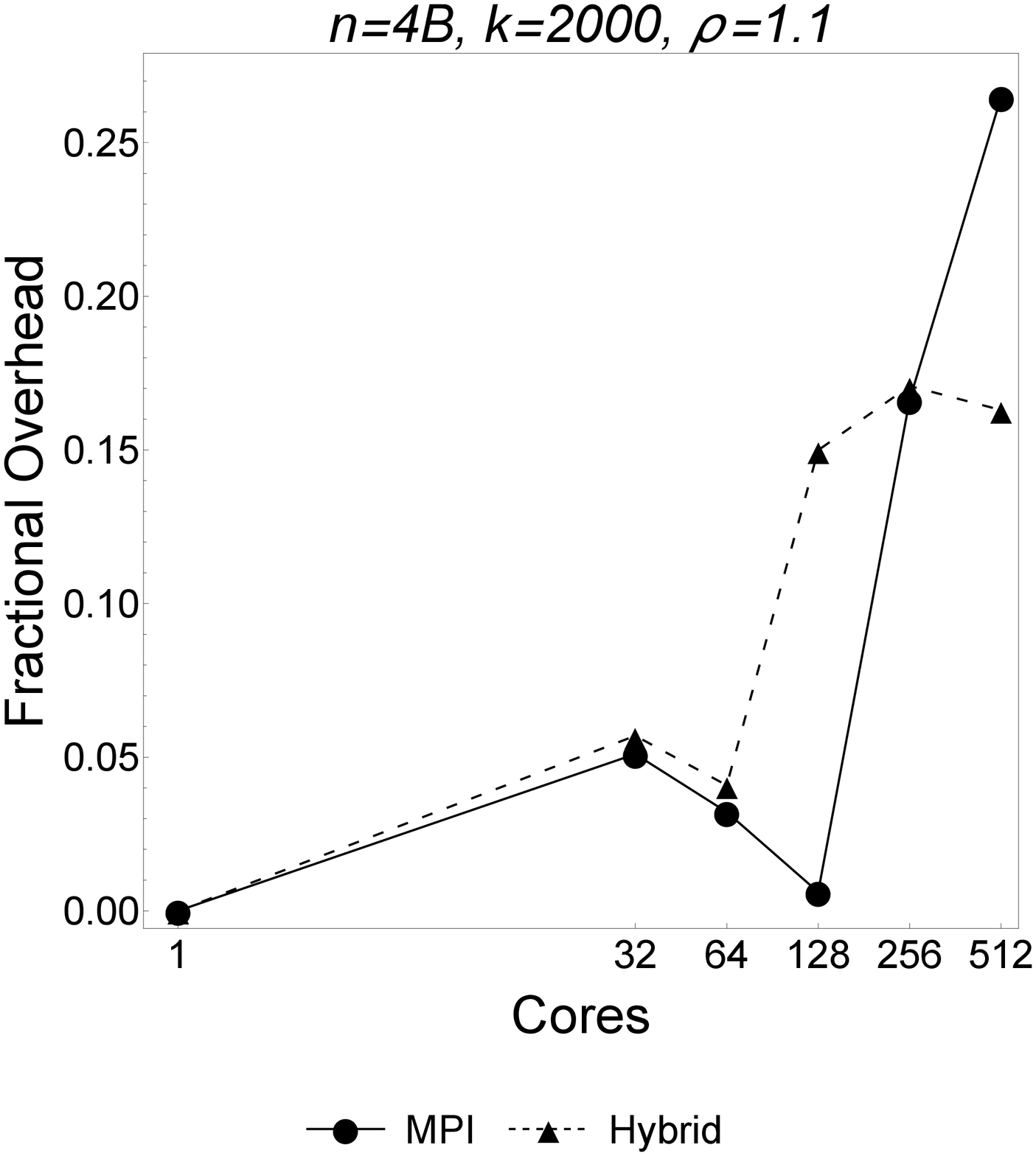}
           \label{mpi_hybrid_comparison_fo_small}
        } &

     \subfloat[Big stream size]{
           \includegraphics[scale=0.22]{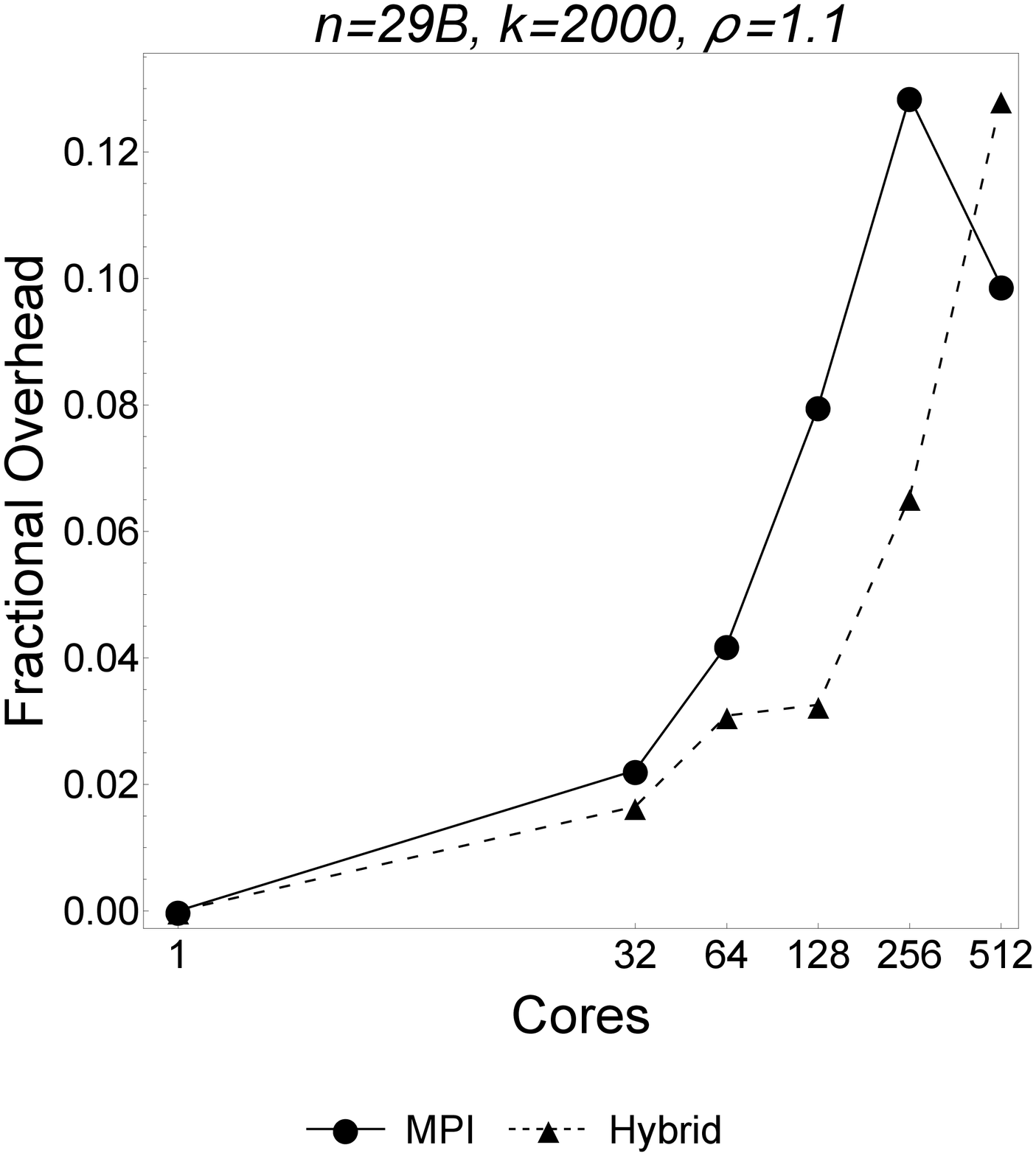}
           \label{mpi_hybrid_comparison_fo_big}
        } \\
        
   \end{tabular}
 \caption{Performance comparison between pure MPI and MPI/OpenMP parallel versions.} 
 \label{mpi_hybrid_comparison}
\end{figure}

The plots in Figure~\ref{mpi_hybrid_comparison} provide the performance comparison between the MPI and MPI/OpenMP versions of the algorithm. We reported only two cases with different input stream sizes (8 and 29 billion items) with skew 1.1 and fixing 2000 counters. The performance of both versions are comparable. The slight differences on the fractional overhead does not introduce significant differences on the speedup between both versions. This can be explained owing to the fact that the algorithm is compute intensive and the communication overhead can have a relevant impact on the performance when using a high number of cores.  
The MPI version has a poor scalability on 512 cores, with a parallel efficiency which is around 50\%. It is worth recalling here that in this case, the input size (29 billion of items) is not enough to provide good scalability on 512 cores owing to the Amdahl effect. With the hybrid parallel version we can achieve better performance with an efficiency greater than 62\% and in some cases reaching 85\%.

\begin{table} [hbt]
\caption{MPI: in each cell the running time (in seconds) and the speedup are reported.}
\label{mpi_wtime_data}
\centering
\def\arraystretch{1.2}
{\scriptsize
\setlength\tabcolsep{5pt}
\begin{tabular}{|c|c|c|c|c|c|c|c|c|c|c|c|}
\hline
\multirow{2}{*}{Cores}  & \multicolumn{4}{c|}{Varying $n$} & \multicolumn{5}{c|}{Varying $k$} & \multicolumn{2}{c|}{Varying $\rho$} \\
 & \text{4B} & \text{8B} & \text{16B} & \text{29B} & 500 & 1000 & 2000 & 4000 & 8000 & 1.1 & 1.8 \\
\hline
\multirow{2}{*}{1} & 122.24 & 238.96 & 481.52 & 874.88 & 953.10 & 892.38 & 874.88 & 935.06 & 976.47 & 874.88 & 689.42 \\
 & 1 & 1 & 1 & 1 & 1 & 1 & 1 & 1 & 1 & 1 & 1 \\
\hline
\multirow{2}{*}{32} & 4.79 & 9.58 & 19.20 & 34.94 & 39.57 & 34.75 & 34.94 & 36.53 & 38.58 & 34.94 & 27.14 \\
 & 25.48 & 24.92 & 25.07 & 25.04 & 24.08 & 25.67 & 25.04 & 25.59 & 25.30 & 25.04 & 25.40 \\
\hline
\multirow{2}{*}{64} & 2.57 & 6.45 & 9.87 & 17.36 & 19.10 & 17.67 & 17.36 & 18.38 & 19.59 & 17.36 & 13.86 \\
 & 47.54 & 37.01 & 48.76 & 50.38 & 49.89 & 50.49 & 50.38 & 50.85 & 49.82 & 50.38 & 49.72 \\
\hline
\multirow{2}{*}{128} & 1.26 & 2.68 & 5.04 & 10.44 & 9.90 & 9.17 & 10.44 & 9.67 & 14.04 & 10.44 & 7.16 \\
 & 96.99 & 88.86 & 95.42 & 83.79  & 96.23 & 97.29 & 83.79 & 96.63 & 69.54 & 83.79 & 96.15 \\
\hline
\multirow{2}{*}{256} & 0.73 & 1.98 & 3.67 & 6.05 & 7.04 & 6.64 & 6.05 & 6.96 & 7.17 & 6.05 & 3.73 \\
 & 167.39 & 120.46 & 131.00 & 144.65  & 135.29 & 134.21 & 144.65 & 134.17 & 136.06 & 144.65 & 184.41 \\
\hline
\multirow{2}{*}{512} & 0.44 & 1.02 & 1.90 & 3.35 & 3.57 & 3.41 & 3.35 & 3.48 & 3.65 & 3.35 & 2.12 \\
 & 272.76 & 232.96 & 253.12 & 261.39 & 266.54 & 261.51 & 261.39 & 268.40 & 267.22 & 261.39 & 323.69 \\
\hline
\end{tabular}
}
\end{table}

\begin{table} [hbt]
 \caption{MPI/OpenMP: in each cell the running time (in seconds) and the speedup are reported.}
 \label{hybrid_wtime_data}
 \centering
 \def\arraystretch{1.2}
{\scriptsize
\setlength\tabcolsep{5pt}
\begin{tabular}{|c|c|c|c|c|c|c|c|c|c|c|c|}
\hline
\multirow{2}{*}{}   & \multicolumn{4}{c|}{Varying $n$} & \multicolumn{5}{c|}{Varying $k$} & \multicolumn{2}{c|}{Varying $\rho$} \\
 & \text{4B}  & \text{8B} & \text{16B} & \text{29B} & 500 & 1000 & 2000 & 4000 & 8000 & 1.1 & 1.8 \\
\hline
\multirow{2}{*}{1} & 119.53 & 238.32 & 479.27 & 869.83 & 1148.35 & 881.50 & 869.83 & 1068.89 & 992.36 & 869.83 & 730.49 \\
 & 1 & 1 & 1 & 1 & 1 & 1 & 1 & 1 & 1 & 1 & 1 \\
\hline
\multirow{2}{*}{32} & 4.81 & 9.62 & 19.09 & 34.47 & 36.72 & 34.73 & 34.47  & 36.20 & 38.08  &  34.47 & 26.43 \\
 & 24.82 & 24.77  & 25.09 & 25.24 & 31.26 & 25.37 & 25.24 & 29.51 & 26.05  &  25.24 & 27.62 \\
\hline
\multirow{2}{*}{64} & 2.55 & 4.92 & 9.67 & 17.38 & 18.68 & 19.70 & 17.38  & 18.37 & 19.40  &  17.38 & 13.42 \\
 & 46.76 & 48.40  & 49.51 & 50.05 & 61.46 & 44.73 & 50.05 & 58.16 & 51.12   &  50.05 & 54.39 \\
\hline
\multirow{2}{*}{128} & 1.28 & 2.45 & 5.45 & 9.26 & 9.48 & 9.90 & 9.26   & 10.34 & 10.29  &   9.26 & 7.59 \\
 & 93.06 & 97.13 & 87.78 & 93.94 & 121.07 & 88.96 & 93.94 & 103.34 & 96.34  &   93.94 & 96.16 \\
\hline
\multirow{2}{*}{256} & 0.72 & 1.28 & 2.79 & 4.87 & 5.33 &   5.01 & 4.87   & 5.21 & 5.54 &   4.87 & 3.68 \\
 & 165.25 & 184.92 & 171.38 & 178.49 & 215.10 & 175.62 & 178.49 & 205.04 & 178.83 &   178.49 & 198.45 \\
\hline
\multirow{2}{*}{512} & 0.36 & 0.67 & 1.46 & 2.40 & 2.60 & 2.46 & 2.40 & 2.53 & 2.86  &   2.40 & 1.92 \\
 & 325.83 & 353.46 & 326.86 & 363.11 & 440.08 & 357.97 & 363.11 & 422.02 & 346.61 &   363.11 & 380.10 \\
 \hline
\end{tabular}
}
\end{table}

\subsection{Evaluation of the OpenMP version on Intel Phi accelerator}
\begin{figure}[hbt]
  \centering
  \begin{tabular}{ccc}
     \subfloat[Running time]{
           \includegraphics[scale=0.16]{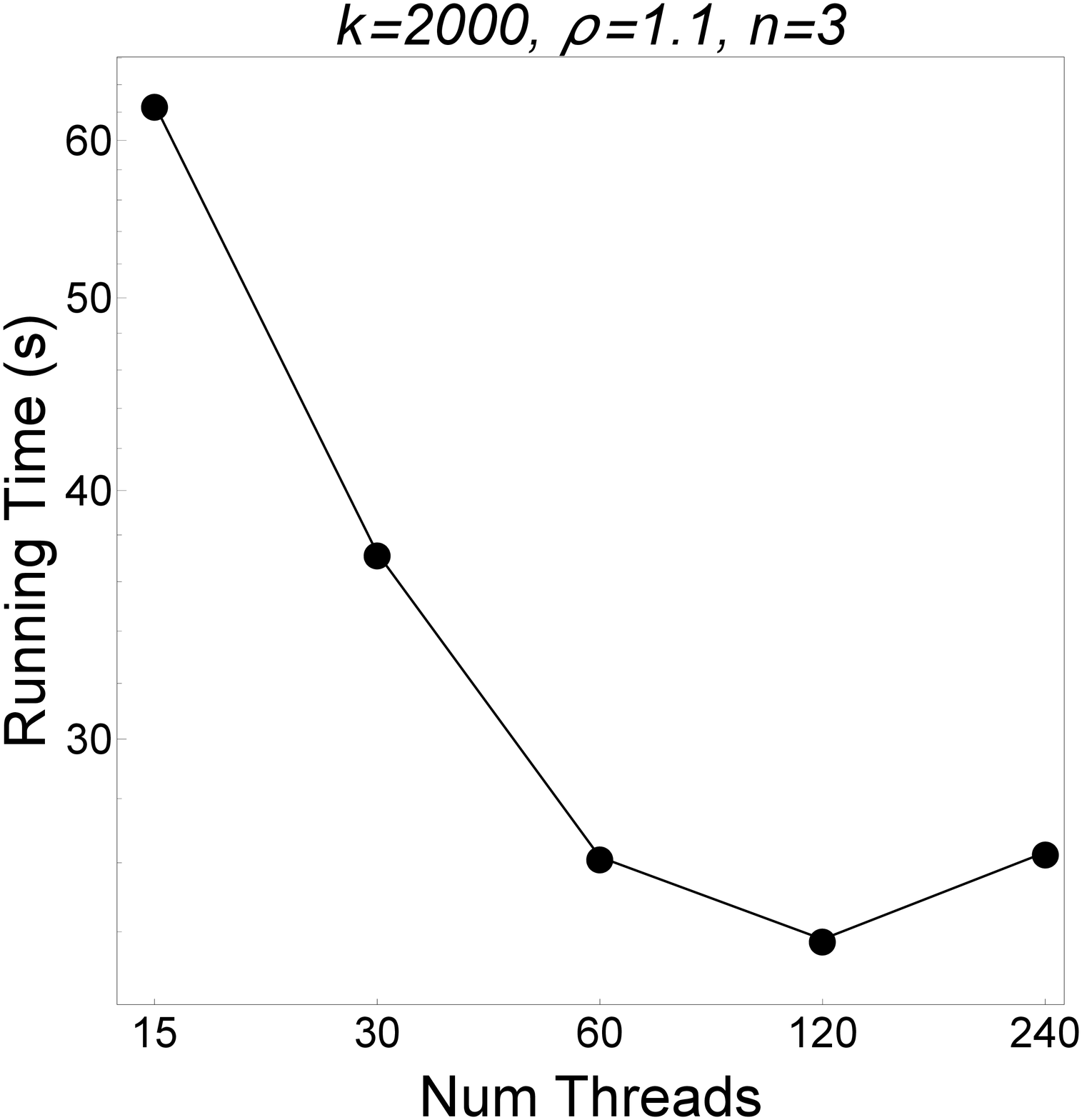}
           \label{xeon_phi_single_rt}
        } &

     \subfloat[Fractional overhead]{
           \includegraphics[scale=0.16]{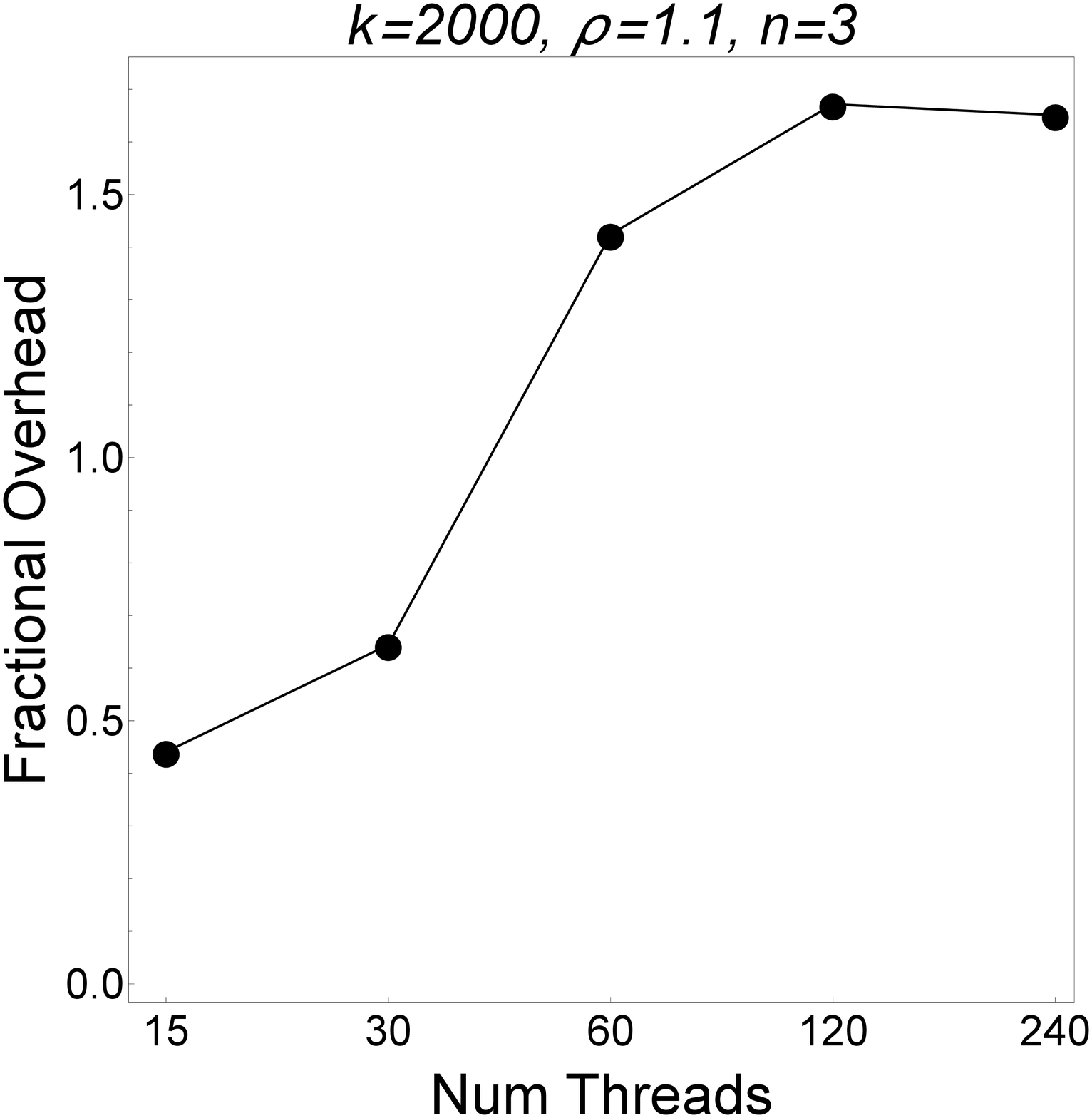}
           \label{xeon_phi_single_fo}
        } &
        
      \subfloat[Average relative error]{
           \includegraphics[scale=0.16]{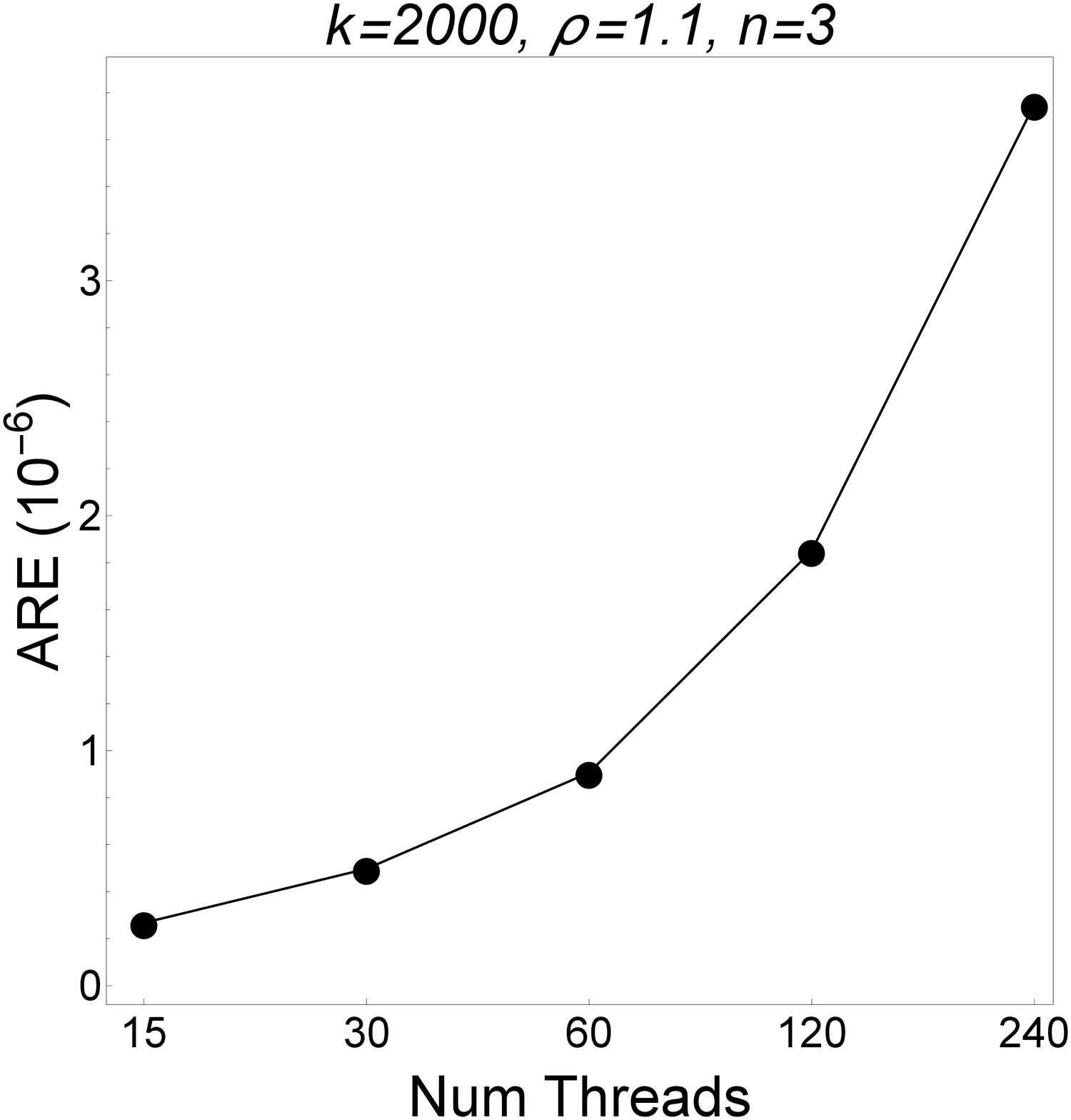}
           \label{xeon_phi_single_are}
        }
   \end{tabular}
 \caption{Performance evaluation on one single Intel Phi accelerator using the OpenMP version.} 
 \label{xeon_phi_single}
\end{figure}

In the third experiment we determined the best configuration of our code on one single Intel Phi accelerator e.g. we measured the performance varying the number of threads on a single Intel Phi accelerator. For this experiment we used an input dataset with 3 billion items with skew 1.1 and 2000 counters. The reason for limiting the dataset size to 3 billion of items is that each Intel Phi accelerator is equipped with 16 GB of RAM and we could not use a bigger input dataset. The results reported in Figure~\ref{xeon_phi_single} reveal that the best performance on a single Intel Phi is obtained by using 120 OpenMP threads, hence the code exploits the Intel Phi accelerator, at the best, with 2 hardware threads (instead of 4).

\subsection{Comparison between Xeon and MIC processors}
In the fourth experiment we evaluated the scalability of the MPI/OpenMP parallel version executed on Intel Phi accelerators instantiating 120 OpenMP threads per MPI process and binding each MPI process to one Intel Phi accelerator; the results were compared to those obtained using the MPI/OpenMP on the Intel Xeon processors instantiating 8 OpenMP threads per MPI process and binding each MPI process to one Xeon processor. We have compared the Phi and Xeon performance varying the number of counters among 500, 1000, 2000, 4000, and 8000 units using an input dataset made of 3 billion items with skew 1.1. Finally, we have also analyzed the behaviour varying the dataset skew between 1.1 and 1.8 with an input dataset of 3 billion items and using 2000 counters.

\begin{figure}[hbt]
  \centering
  \begin{tabular}{ccc}
     \subfloat[Running time]{
           \includegraphics[scale=0.22]{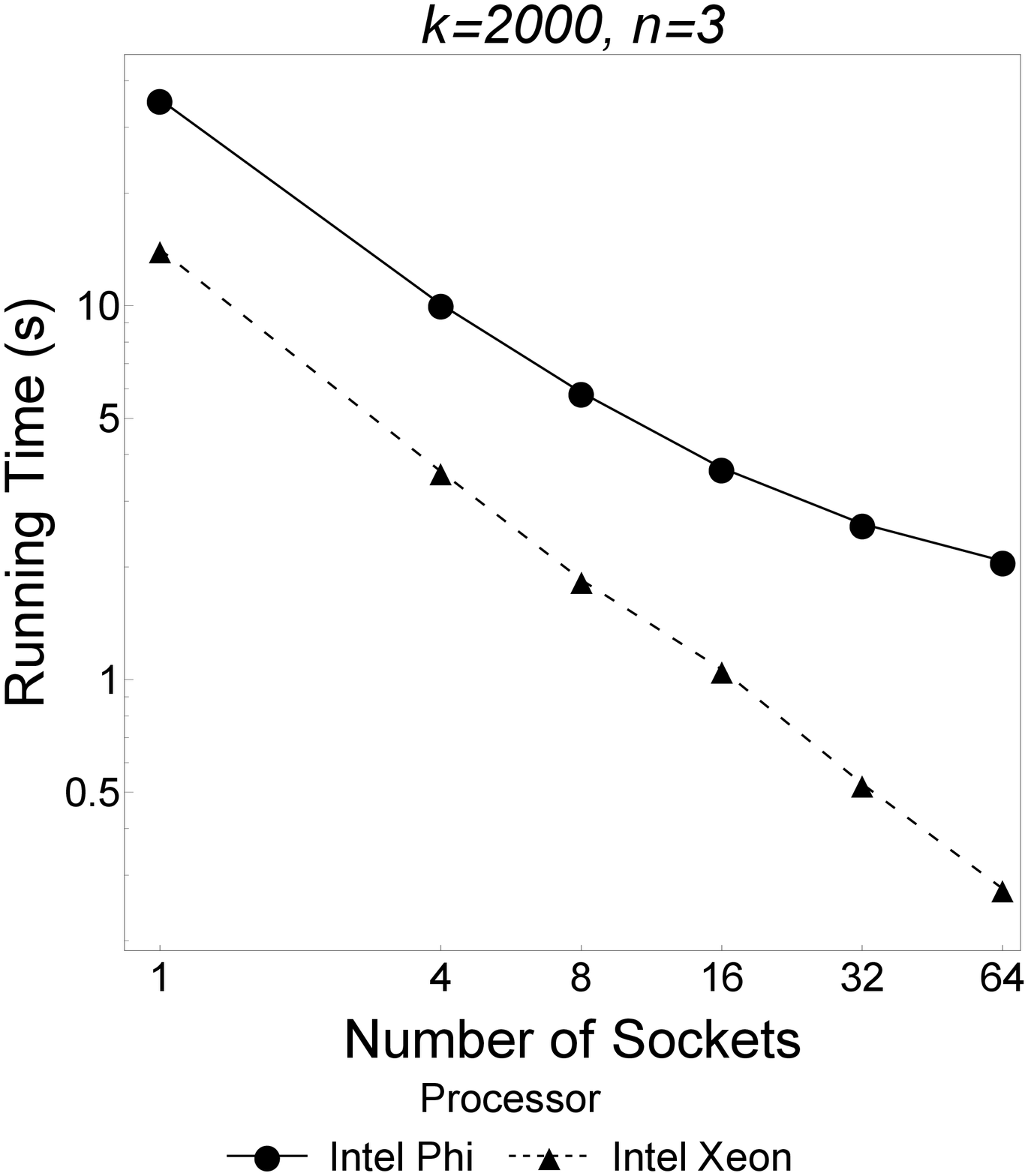}
           \label{xeon_mic_comparison_k}
        } &

     \subfloat[Fractional overhead]{
           \includegraphics[scale=0.22]{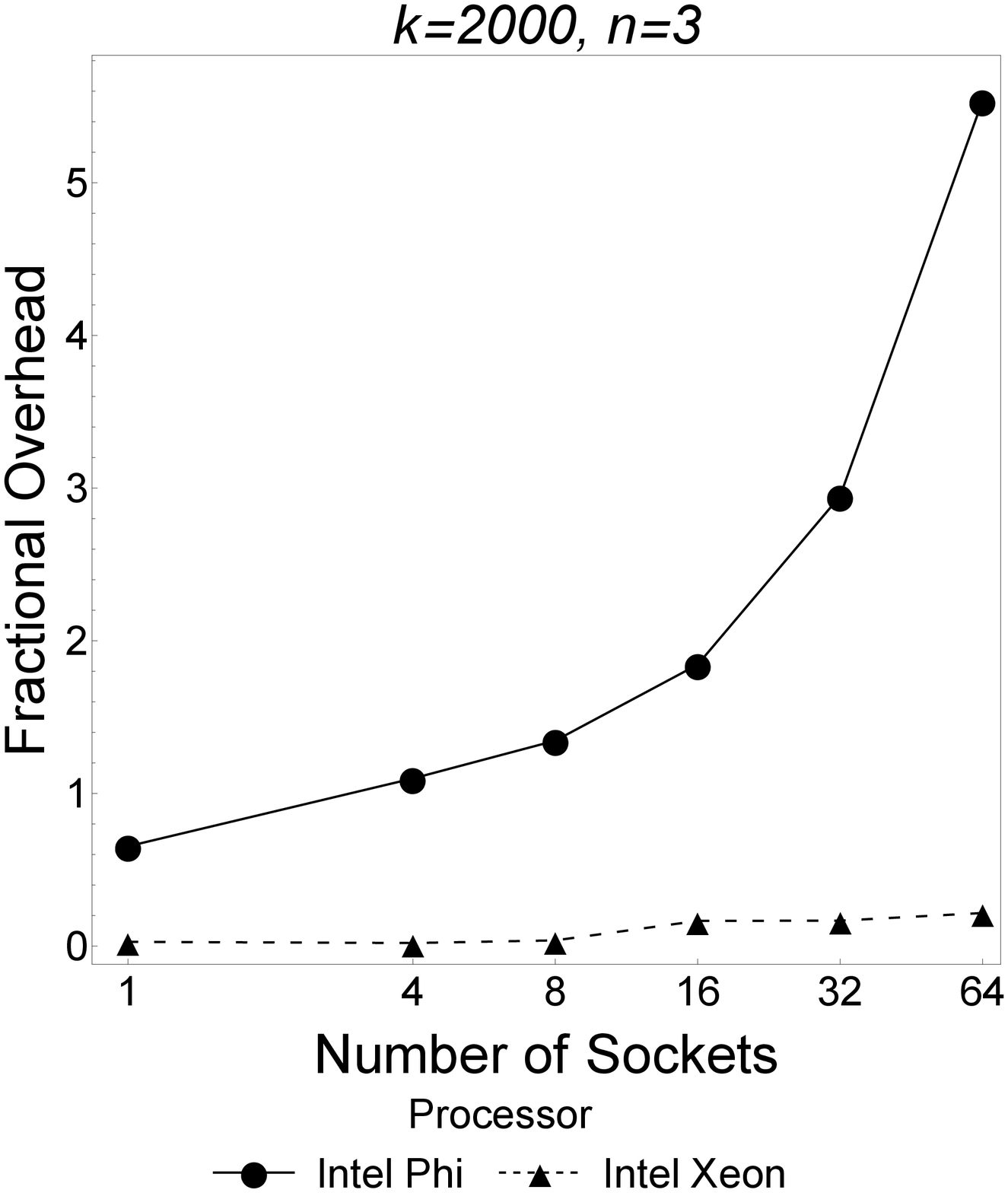}
           \label{xeon_mic_comparison_n}
        } &
        
      \subfloat[Average relative error]{
           \includegraphics[scale=0.22]{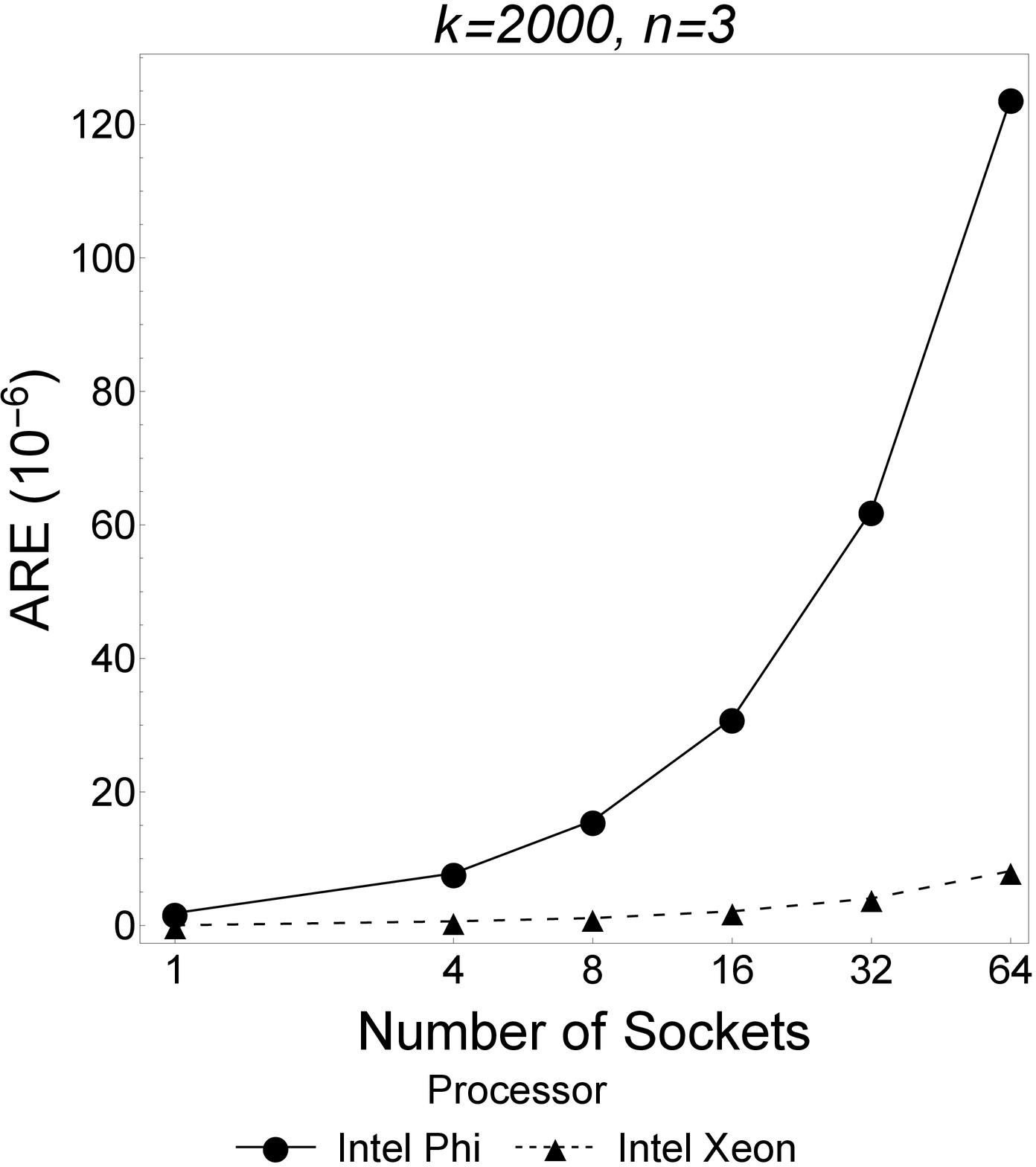}
           \label{xeon_mic_comparison_ro}
        }
   \end{tabular}
 \caption{Performance comparison between the Intel Xeon and Intel Phi using MPI/OpenMP parallel version.} 
 \label{xeon_mic_comparison}
\end{figure}

The MPI/OpenMP version has been executed on Intel Phi offloading to the MIC accelerator the computational part of the code and the interprocess communication, while the I/O operations are executed on the CPU. Figure~\ref{xeon_mic_comparison} shows that execution on Intel Phi accelerator did not provide any advantage with regard to the Intel Xeon processor. The motivation is twofold: our algorithm uses an unordered and unpredictable pattern for memory accesses during the items' frequency update owing to the use of hash tables and this heavily limits the exploitation of the 512-bit wide SIMD vector unit included in the Intel Phi accelerator; in addition, due to the non-contiguous access to memory, the algorithm does not exhibit any data locality limiting the exploitation of the cache memory hierarchy. Hence, even if the algorithm exposes a high level of data parallelism, it does not fit well the Intel Phi accelerator architecture, making the use of the traditional Xeon processor more suitable for good performance. 

\section{Conclusions}
\vspace{-2pt}
\label{conclusions}

We have presented a shared-memory version of our Parallel Space Saving algorithm, and studied its behavior with regard to accuracy and performance on multi and many-core processors, including the Intel Xeon and Phi. In particular, we carried out several experiments and reported extensive experimental results, including a comparison of a hybrid MPI/OpenMP version against a pure MPI based version. The results demonstrated that the MPI/OpenMP parallel version of the algorithm significantly improve the parallel speedup and efficiency due to the use of the shared-memory approach reduces the communication overhead introduced by the parallel reduction. Moreover we have also experimented that for its characteristics, the Intel Phi accelerator is not suitable for this algorithm since the algorithm exhibits a highly limited data locality and the non-contiguous memory access limit the exploitation of the cache hierarchy.

\ack The authors would like to thank CINECA for granting the access to the Galileo supercomputer machine through grant IsC40\_PFI HP10C84WLO.

\bibliography{pss}
\bibliographystyle{wileyj}

\end{document}